\documentclass[graybox,vecphys]{svmult}

\usepackage{comment}
\usepackage[T1]{fontenc}
\usepackage{bm,marvosym,amsmath,empheq}

\usepackage{type1cm}        
\usepackage{graphicx}        
\usepackage{multicol}        
\usepackage[bottom]{footmisc}

\usepackage{newtxtext}       %
\usepackage[varvw]{newtxmath}       

\usepackage[square,numbers,sort]{natbib}

\usepackage[colorlinks,citecolor=blue]{hyperref}
\usepackage{bookmark}

\makeatletter
\def\toclevel@title{-1}
\def\toclevel@author{0}
\makeatother

\begin{document}

\title*{Variational Acoustics}
\author{Md Mahmudur Rahman\protect\texorpdfstring{\orcidID{0009-0004-6612-4644}}{} and\protect\texorpdfstring{\\}{} Ivan C.\ Christov\protect\texorpdfstring{\orcidID{0000-0001-8531-0531}}{}}
\institute{Md Mahmudur Rahman \at School of Mechanical Engineering, Purdue University, West Lafayette, Indiana 47907, USA \email{rahma165@purdue.edu}
\and Ivan C.\ Christov (\Letter) \at School of Mechanical Engineering, Purdue University, West Lafayette, Indiana 47907, USA \email{christov@purdue.edu}}
%
%
\maketitle


\abstract{Many variational principles for acoustics have appeared in the literature, often introduced in an \textit{ad hoc} manner by picking and choosing terms in a Lagrangian density to yield a desired governing partial differential equation. In this chapter, we show that such guesswork is unnecessary, as the governing equations of acoustics can be derived systematically from the primitive Lagrangians of classical continuum mechanics. Furthermore, we note that using Eulerian coordinates is not the only way to derive variational principles in acoustics. To this end, we review the construction of an action in the Lagrangian frame for compressible nondissipative fluids. We explore the consequences of material relabeling symmetry and the associated pseudomomentum balance, together with the boundary and jump conditions it implies. Via perturbation expansions, we show how to recover the equations of linear acoustics from each variational principle, the material frame yielding in addition the second-order acoustic energy balance. We show how weakly nonlinear approximations with variational structure emerge in each frame, in both cases for a general barotropic fluid, with the nonlinearity carried by the fluid's parameter $B/A$ and the perfect-gas results recovered as a special case.}


\section{Introduction}

A survey of the literature suggests that variational formulations of acoustics have often been introduced in an \emph{ad hoc} manner \cite{Morse1968TheoreticalAcoustics,Rasmussen2008AnalyticalDissipation,Rasmussen2010AnalyticalDissipation,Christov2015CorrigendumPropagation}. What we mean is that, in such approaches, the Lagrangian density for the variational principle is postulated based on the desire to arrive at a particular governing partial differential equation (PDE). Consequently, many variational principles for acoustics have appeared in the literature. Although such formulations can be useful, they are often reverse-engineered from a wave equation and suitable approximations, rather than derived from a more fundamental mechanistic framework. Thus, though useful for certain calculations, these approaches inherently obscure the underlying variational structure. 

Since acoustics is ultimately a classical continuum theory, it is natural to ask whether one may instead begin with the standard Lagrangian framework of continuum mechanics and derive the acoustic equations systematically. In this chapter, we show how some familiar \emph{ad hoc} Lagrangians of acoustics can be systematically derived from the primitive Lagrangian density. This procedure can be performed in either the Eulerian or Lagrangian frame. We consider both cases in parallel.

Frisch et al.~\cite{Frisch2017ADynamics} argue that Hermann Hankel (see \cite{Villone2017Hermann1861} for an English translation) was the first to introduce a variational principle for nondissipative \emph{compressible} fluids in 1861, from both viewpoints by ``constant juggling with Eulerian and Lagrangian coordinates.'' Hargreaves \cite{Hargreaves1908APotential} realized that certain connections exist between integrals of the pressure, the corresponding Euler--Lagrange equations that extremize such integrals, and the equations of fluid motion in the Eulerian frame. Interestingly, he even formulated a variational principle for the linear acoustic wave equation at the end of his paper \cite{Hargreaves1908APotential}. However, it appears that most modern treatments now begin with Bateman's 1929 work \cite{Bateman1929NotesProblems}, which builds on Hargreaves' observations but still focuses largely on determining Lagrangian densities that yield certain governing equations, essentially the inverse (rather than direct) approach. Eckart's 1938 work \cite{Eckart1938TheMedia} on constraining the Lagrangian to enforce continuity equations briefly considers the nondissipative fluid. But Herivel \cite{Herivel1955ThePrinciple} appears to have been the first to construct, in 1955, a modern variational principle for nondissipative fluids explicitly in the Lagrangian description and connecting to Eckart's Eulerian description by a change of variables. Eckart's more authoritative treatment of the adiabatic motion of a compressible fluid in Lagrangian coordinates came in 1960~\cite{Eckart1960VariationHydrodynamics}. Then, in 1968, Seliger and Whitham \cite{Seliger1968VariationalMechanics} unified the treatment of various variational problems in a modern form by a direct approach. Salmon's 1988 review \cite{Salmon1988HamiltonianMechanics} gives a comprehensive treatment of variational formulations and their history for fluids in both the Lagrangian and Eulerian descriptions and also clarifies the correspondence between the two descriptions.

Variational principles appear in acoustics textbooks as early as 1968 in Morse and Ingard's \textit{Theoretical Acoustics}~\cite[Sec.~6.2]{Morse1968TheoreticalAcoustics} (presumably building on the earlier field-theoretic exposition of Morse and Feshbach \cite[Chap.~3]{MorseFeshbach1953}), but they do not mention Hargreaves' work \cite{Hargreaves1908APotential}. Pierce \cite{Pierce1993VariationalScattering} also discussed variational principles (constructed \textit{a posteriori}) in the context of elastic media and acoustic scattering. A few of these topics made it into his influential textbook \textit{Acoustics}~\cite{Pierce2019Acoustics}, but primarily as a tool for computing.

Meanwhile, variational acoustics from the point of view of the Lagrangian frame has not been popular in the literature for many years. Nevertheless, an interesting aspect of the calculations in the Lagrangian frame is their connection to the framework of configurational mechanics, which elegantly unifies aspects of mechanics~\cite{Fried2005TheBalance}. 
From a historical perspective, one may trace important precursors of the configurational viewpoint to the work of Peierls~\cite{Peierls1976TheMedium}, who clarified aspects of the Abraham--Minkowski debate concerning momentum and wave momentum. In that setting, Peierls discussed the concept of \emph{pseudomomentum} and identified an associated balance law linked to transformations of the material medium, although his derivation was carried out within the Poisson bracket formalism of classical mechanics.  

The modern approach begins with Eshelby's formulation of the elastic energy--momentum tensor~\cite{Eshelby1951TheSingularity,Eshelby1975TheTensor} and uses symmetry arguments to obtain driving forces on defects and forces acting on inhomogeneities. Related developments may be found in~\cite{Gupta2007AnSpace,Markenscoff2006EshelbyIntegrals,Markenscoff2023LossDislocation}. Configurational, or material forces also play an important role in applications such as shock-front propagation~\cite{Maugin1997DrivingCrystals} and phase transformations~\cite{Abeyaratne2006EvolutionTransitions}. The contemporaneous works of Rice~\cite{Rice1968ACracks} and Cherepanov~\cite{Cherepanov1967CrackMedia} demonstrated that a contour integral surrounding a notch or crack tip is invariant with respect to the chosen path. Later, Go\l{}\k{e}biewska Herrmann reformulated elasticity in a Lagrangian framework and showed that the $J$-integral may be interpreted as the pseudomomentum flux of the system~\cite{Golebiewska-Herrmann1983OnMechanics}. More generally, we think of configurational forces as arising from the breaking of the material relabeling symmetry of a system, whereas, when such material symmetry is present, the associated pseudomomentum is conserved. In this way, physical phenomena that appear, at first sight, to be entirely unrelated, may nevertheless emerge from the same underlying mathematical structure~\cite{Singh2021Pseudomomentum:Consequences}. 

The Lagrangian-frame treatment draws on concepts of material coordinates, covariant derivatives, and variational identities on a manifold; thus, it may appear ``heavier'' than the Eulerian treatment that precedes it. The investment in introducing and using this machinery is justified by what it yields naturally and without additional hypotheses. For example, when applied to fluid mechanics, we learn that vorticity conservation is just a consequence of material relabeling symmetry~\cite{Singh2021Pseudomomentum:Consequences,Badin2018VariationalDynamics}. Other quantities, such as helicity conservation and the Cauchy--Weber integral, may likewise be obtained from the bulk balance of pseudomomentum~\cite{Singh2021Pseudomomentum:Consequences}. Related ideas have also proved useful in interfacial mechanics. For example, configurational forces have been used to study the interfacial properties of elastomers, including problems of wettability~\cite{Masurel2019ElastocapillaryDisclination}. They may also be used to determine the shape of a sessile drop on a substrate, in which case they lead to a generalized Young--Laplace equation~\cite{Lubarda2012ConfigurationalSubstrate}.

In this chapter, we focus on compressible nondissipative fluid flows, as in classical acoustics. We first consider the formulation of the variation principle in the Eulerian frame, making use of Eckart's ``trick'' \cite{Eckart1938TheMedia,Eckart1960VariationHydrodynamics}, introducing a Lagrange multiplier to enforce the continuity equation (Sec.~\ref{sec:Eulerian}). Then, we apply the formalism of Singh and Hanna \cite{Singh2021Pseudomomentum:Consequences} to compressible fluid flow and derive the associated balance laws in their most general form directly from the symmetries of the Lagrangian density (Sec.~\ref{sec:configurational}). In each frame, we carry the perturbation expansion of the Lagrangian density to cubic order, obtaining weakly nonlinear wave equations that hold for barotropic fluids; the material-frame version of this calculation appears to be new. We conclude the discussion in Sec.~\ref{sec:conclusion}.

\section{Variational principle for acoustics in the Eulerian description}
\label{sec:Eulerian}

Let us first discuss the ``standard'' approach to variational principles of acoustics, developed using the Eulerian description of continuum mechanics. Specifically, we shall revisit several aspects of the contributions of Eckart \cite{Eckart1960VariationHydrodynamics} and Seliger and Whitham \cite{Seliger1968VariationalMechanics} and specialize to the context of acoustics.

\subsection{Background}
\label{sec:Eulerian_background}

In acoustics, it is of interest to study perturbations about a quiescent equilibrium (``ambient'') state of a compressible fluid (i.e., a ``gas''). Restricting to nondissipative fluid flows, i.e., ones in which viscous forces and thermal transport are negligible, let us denote the quiescent state as $(\rho, p, \bm{v}) = (\rho_0, p_0, \bm{0})$, where $\rho_0$ and $p_0$ are the constant reference density and pressure \cite{Pierce2019Acoustics}. Then, the density, velocity, and pressure fields are expanded, respectively, as 
\begin{subequations}\begin{align}
    \rho
    &=
    \rho_0
    + \rho^{(1)}
    +\text{h.o.t.},\\
    \bm{v}
    &= 
    \bm{0}
    + \bm{v}^{(1)}
    +\text{h.o.t.},\\
    p &= p_0
    +
    p^{(1)}
    +\text{h.o.t.}, 
\end{align}\end{subequations}
where ``h.o.t.'' stands for ``higher-order terms,'' and there is an implied asymptotic ordering, such as $\rho^{(1)} \ll \rho_0$, etc.\ \cite{Pierce2019Acoustics}. 

Since the flow is nondissipative, entropy is conserved along particle paths. Further, since we perturb about a spatially uniform quiescent state, the entropy is uniform throughout (i.e., the flow is \emph{homentropic}). The terminology varies in the literature, thus we clarify that we shall use these terms in the following hierarchy: \emph{nondissipative} refers to the neglect of viscous forces and thermal transport; \emph{isentropic} to the resulting conservation of entropy along particle paths, $\mathrm{d}s/\mathrm{d}t=0$; and \emph{homentropic} to the stronger statement that $s$ is in addition uniform in space. Homentropic implies isentropic, so thermodynamic identities derived for an isentropic process hold \textit{a fortiori} for a homentropic one. But, importantly, it is the additional spatial uniformity that permits a \emph{barotropic} closure $p=p(\rho)$, with a single function $p(\cdot)$ valid throughout the fluid. Unless stated otherwise, every result below assumes a nondissipative, homentropic (hence, barotropic) fluid flow in the absence of body forces. The perfect-gas relations \eqref{eq:isentropic_rels} are a further specialization, used only in Secs.~\ref{sec:different_Ls} and~\ref{sec:peturbation_Eulerian} to obtain the closed-form Lagrangian density \eqref{eq:final lagrangian density}. Everything thereafter, in both Secs.~\ref{sec:Eulerian} and~\ref{sec:configurational}, holds for a general barotropic fluid, with the perfect gas recovered as a special case.

With this in mind, $p_0=p(\rho_0)$, and a Taylor expansion of $p(\rho)$ about the ambient state introduces two equivalent measures of stiffness: the isentropic \emph{compressibility} $\kappa_0=[\rho_0(\mathrm{d}p/\mathrm{d}\rho)|_{\rho=\rho_0}]^{-1}$, in which Morse and Ingard \cite[p.~230]{Morse1968TheoreticalAcoustics} and Pierce \cite{Pierce2019Acoustics} write the linear theory, and the sound speed $c_0^2=(\mathrm{d}p/\mathrm{d}\rho)|_{\rho=\rho_0}=1/(\kappa_0\rho_0)$, which we use hereafter.

Now, a standard exercise, using the conservation of mass and balance of linear momentum equations of a nondissipative compressible fluid, yields 
\begin{subequations}\label{eq:morse_ingard_linear_acoustics}\begin{align}
    \frac{\partial \rho^{(1)}}{\partial t} &= - \rho_0 \nabla \cdot \bm{v}^{(1)}, \label{eq:rho1_v1}\\
    \rho_0 \frac{\partial \bm{v}^{(1)}}{\partial t} &= - \nabla p^{(1)},
    \label{eq:v1_p1}\\
    \rho^{(1)} & = \kappa_0 \rho_0 p^{(1)},
\end{align}\end{subequations} 
where the last relation is the linearized barotropic equation of state, equivalently $p^{(1)} = c_0^2\rho^{(1)}$. 
This calculation is available in almost every introductory textbook \cite{Morse1968TheoreticalAcoustics,Fetter2003TheoreticalContinua,Blackstock2000FundamentalsAcoustics,Pierce2019Acoustics,Gonzalez2001AMechanics}. 
Conventionally, taking the curl of \eqref{eq:v1_p1} lets us infer that $\partial \bm{\omega}^{(1)}/\partial t = \bm{0}$, where $\bm{\omega}^{(1)} = \nabla\times\bm{v}^{(1)}$ is the vorticity. Thus, a flow free of initial vorticity ($\bm{\omega}^{(1)} = \nabla\times\bm{v}^{(1)} = \bm{0}$ at $t=0$) remains free of vorticity (for all $t>0$), at least \emph{at first order}. Such an \emph{irrotational} flow allows the introduction of the \emph{acoustic potential} $\Psi$ via $\bm{v} = \bm{0} + \bm{v}^{(1)} + \cdots = - \nabla \Psi$ \cite{Pierce2019Acoustics}.\footnote{The fully-nonlinear counterpart of this statement is Kelvin's circulation theorem (for a barotropic nondissipative fluid), derived in Sec.~\ref{sec:configurational} below. It shows that an initially irrotational flow remains such, which is why we can introduce a velocity potential in the nonlinear problem, reusing the symbol $\Psi$.} Then, it is straightforward to combine equations \eqref{eq:morse_ingard_linear_acoustics} into
\begin{equation} 
    \frac{1}{c_0^2}\frac{\partial^2 \Psi}{\partial t^2} - \nabla^2 \Psi = 0,
    \label{eq:morse_ingard_wave_equation} 
\end{equation}
where $c_0$ is the sound speed defined above. Equation~\eqref{eq:morse_ingard_wave_equation} is the \emph{linear} acoustic wave equation in terms of the acoustic potential. This PDE is the entry point into the field of acoustics.

Next, Morse and Ingard \cite[p.~248]{Morse1968TheoreticalAcoustics} discuss a variational principle for linear acoustics. In an essentially \emph{ad hoc} manner, they propose\footnote{Note that Hargreaves \cite{Hargreaves1908APotential} had already obtained, 60 years earlier, the Lagrangian density~\eqref{eq:morse_ingard_lagrangian_component}, up to an overall sign, though not by the calculus of variations. Instead, he assumed the wave equation~\eqref{eq:morse_ingard_wave_equation} holds and used it to reduce his ``kinetic potential'' to surface integrals.} a Lagrangian density $\mathcal{L}$:
\begin{equation}
    \mathcal{L}_\mathrm{lin}
    =
    \frac{1}{2}\rho_0
    \left[
    c_0^{-2} \left(\frac{\partial\Psi}{\partial t}\right)^2
    -|\nabla\Psi|^2
    \right].
    \label{eq:morse_ingard_lagrangian_component}
\end{equation}
Requiring the action $A = \int \mathrm{d}t \int \mathrm{d}\bm{x} \, \mathcal{L}$ be stationary for arbitrary $\delta \Psi$  vanishing on the boundary leads to the Euler--Lagrange equation \cite{Goldstein2002ClassicalMechanics,Fetter2003TheoreticalContinua}:
\begin{equation}
    \frac{\partial}{\partial t}\left(\frac{\partial \mathcal{L}}{\partial (\partial \Psi/\partial t)}\right)
    + \nabla\cdot\left(\frac{\partial \mathcal{L}}{\partial (\nabla\Psi)}\right)
    -\frac{\partial \mathcal{L}}{\partial \Psi}
    =0.
\label{eq:EL operator}
\end{equation}
Substituting $\mathcal{L}= \mathcal{L}_\mathrm{lin}$ from \eqref{eq:morse_ingard_lagrangian_component}  into \eqref{eq:EL operator} indeed gives the wave equation~\eqref{eq:morse_ingard_wave_equation} as desired.

Then, as a starting point for \emph{nonlinear} acoustics, Morse and Ingard~\cite[p.~249]{Morse1968TheoreticalAcoustics} proposed including higher-order terms in \eqref{eq:morse_ingard_lagrangian_component}, namely cubic products of $\Psi$ and its derivatives, specifically as
\begin{equation}
    \mathcal{L}_{\mathrm{MI}}
    =
    \frac{1}{2}\rho_0
    \left[
    1 + c_0^{-2} \left(\frac{\partial\Psi}{\partial t}\right)
    \right]
    \left[
    c_0^{-2} \left(\frac{\partial\Psi}{\partial t}\right)^2
    -|\nabla\Psi|^2
    \right].
\label{eq:morse_ingard_nonlinear_lagrangian}
\end{equation}
However, Christov and Jordan~\cite{Christov2015CorrectionsGases} showed that (i) the nonlinear acoustic wave equation quoted in \cite{Morse1968TheoreticalAcoustics} does not follow from $\mathcal{L}_{\mathrm{MI}}$, and (ii) the Lagrangian density for nonlinear acoustics ought to be 
\begin{equation}
    \mathcal{L}_{\mathrm{corr}}
    =
    \frac{1}{2}\rho_{0}
    \left[
    c_{0}^{-2}\left(\frac{\partial\Psi}{\partial t}\right)^{2}
    -|\nabla\Psi|^{2}
    -\frac{1}{3}(\gamma-2)c_{0}^{-4}\left(\frac{\partial\Psi}{\partial t}\right)^{3}
    -c_{0}^{-2}\left(\frac{\partial\Psi}{\partial t}\right)|\nabla\Psi|^{2}
    \right],
    \label{eq:christov_jordan_corrected_lagrangian}
\end{equation}
where $\gamma$ is the so-called adiabatic index (ratio of specific heats of a perfect gas). 

At this point, one ought to ask: What are the underlying \emph{physical principles} leading to \eqref{eq:christov_jordan_corrected_lagrangian}, and therefore \eqref{eq:morse_ingard_lagrangian_component}? Rather than try to infer or guess a suitable $\mathcal{L}$ that gives the governing PDE variational structure, we discuss what variational structure exists \textit{a priori}, based on principles of continuum mechanics, from which $ \mathcal{L}_{\mathrm{corr}}$ and any attendant PDE ought to arise.

\subsection{Lagrange multipliers and Eckart's principle}
\label{sec:Eckart}

Christov and Jordan~\cite{Christov2015CorrectionsGases} suggested that the Lagrangian density \eqref{eq:christov_jordan_corrected_lagrangian} is a Taylor series expansion of a more general Lagrangian density. Thus, let us fill in the steps leading us to the ``more general'' one posited in \cite{Christov2015CorrigendumPropagation}. Let us follow a fundamental approach, similar to that in classical mechanics \cite{Goldstein2002ClassicalMechanics,Fetter2003TheoreticalContinua}. To obviate guesswork in constructing the variational principle, it is necessary to return to primitive variables and not attempt to immediately formulate a variational problem in terms of the acoustic potential. 

Two fields are necessary to describe compressible fluid motions: the density $\rho(\bm{x},t)$ and the velocity $\bm{v}(\bm{x},t)$. Furthermore, $\rho$ and $\bm{v}$ \emph{cannot} vary independently; they are constrained by the continuity (conservation of mass) condition. Following Eckart~\cite{Eckart1938TheMedia,Eckart1960VariationHydrodynamics}, a Lagrangian density that takes this constraint into account is 
\begin{equation}
  \mathcal{L}
  = \underbrace{\frac{1}{2} \rho \lvert \bm{v} \rvert^2}_{\text{kinetic energy density}}
    - \underbrace{\rho\, e(\rho,s)}_{\text{internal energy density}}
    - \;\;\; \underbrace{\Psi\big[ \partial_t \rho + \nabla \cdot (\rho \bm{v}) \big]}_{\text{constraint}} \, ,
  \label{eq:lagrangian_density}
\end{equation}
where $\Psi$ is a \emph{Lagrange multiplier} enforcing the constraint
(continuity equation). With the constraint so enforced, $\rho$, $\bm{v}$, and
$\Psi$ may henceforth be varied independently. Henceforth, we use shorthand notation $\partial_t = \partial/\partial t$. The dependence of the specific internal energy $e$ on the mass density $\rho$ and entropy $s$ is supplied by a thermodynamic equation of state. Since the flow is homentropic, $s$ is a fixed constant throughout, so we write $e(\rho)$ from here on in this section, with total derivatives in $\rho$. Similarly, and unlike incompressible flows, the pressure field $p$ is linked to $\rho$ (and, perhaps, $e$ and/or $s$) via thermodynamics and is therefore not independent. 

Equation~\eqref{eq:lagrangian_density} can be rearranged as
\begin{equation}
  \mathcal{L}
  = \frac{1}{2} \rho \lvert \bm{v} \rvert^2
    - \rho\, e
    - \Psi\,\partial_t\rho
    + \rho\,\bm{v}\cdot\nabla\Psi
    - \nabla\cdot(\Psi\rho\bm{v}).
  \label{eq:lagrangian_density_2}
\end{equation}
Note that the last term is a total divergence, contributing a boundary term that drops out for variations vanishing on the boundary. Also, we may write $-\Psi\partial_t\rho = \rho\partial_t\Psi- \partial_t(\Psi\rho)$. Thus, we can equivalently rewrite \eqref{eq:lagrangian_density_2} as
\begin{equation}
  \mathcal{L}
  = \frac{1}{2} \rho \lvert \bm{v} \rvert^2
    - \rho\, e
    + \rho\partial_t\Psi
    + \rho\bm{v}\cdot\nabla\Psi
    - \partial_t(\Psi\rho),
  \label{eq:lagrangian_density_3}
\end{equation}
where again, the last term contributes a boundary term to the action, and this term vanishes upon variation.
Finally, we obtain the ``simplest'' equivalent form of \eqref{eq:lagrangian_density}:
\begin{equation}
  \mathcal{L}
  = \frac{1}{2} \rho \lvert \bm{v} \rvert^2
    - \rho\, e
    + \rho\partial_t\Psi
    + \rho\bm{v}\cdot\nabla\Psi.
  \label{eq:lagrangian_density_4}
\end{equation}
We term \eqref{eq:lagrangian_density_4} the ``primitive'' Lagrangian density for acoustics. Fetter and Walecka \cite[p.~304]{Fetter2003TheoreticalContinua} discuss several aspects of this calculation but do not provide the full context, leaving the reader to think that $\mathcal{L}$ was guessed rather than derived.

\subsection{Different ways of recasting the primitive Lagrangian density}
\label{sec:different_Ls}

Now, we can explore some consequences of \eqref{eq:lagrangian_density_4}. Since $\bm{v}$ and $\rho$ enter \eqref{eq:lagrangian_density_4} algebraically, stationarity of the action with respect to them reduces to the pointwise vanishing of the corresponding variation of $\mathcal{L}$; only $\Psi$ requires the full operator \eqref{eq:EL operator}.

Variation with respect to $\bm{v}$ yields
\begin{equation}
  \delta_{\bm{v}} \mathcal{L}
  = \rho\bm{v}\cdot\delta\bm{v}
    + \rho\nabla\Psi\cdot\delta\bm{v}
  = \rho(\bm{v} + \nabla\Psi)\cdot\delta\bm{v}.
  \label{eq:variation_L_v}
\end{equation}
Requiring the variation $\delta_{\bm{v}} \mathcal{L}$ vanish for arbitrary $\delta\bm{v}$ leads to
\begin{equation}
      \bm{v} = -\nabla\Psi,
  \label{eq:velocity_potential}
\end{equation}
which leads us to identify $\Psi$ as the familiar acoustic (velocity) potential. We stress that \eqref{eq:velocity_potential} does not \emph{demonstrate} irrotationality; rather, it reveals that the Lagrangian \eqref{eq:lagrangian_density}, constrained by mass conservation alone, can represent \emph{only} irrotational motions.\footnote{Historically (see discussion in, e.g., \cite{Seliger1968VariationalMechanics,Bretherton1970AFluids}), C.~C.~Lin identified this deficiency of Eckart's constraint ``trick.'' Clebsch variables can be introduced through further constraints in \eqref{eq:lagrangian_density} to handle flows that are not irrotational (curl-free) \cite{Eckart1960VariationHydrodynamics,Seliger1968VariationalMechanics,Tong2025FluidMechanics}, but that is not of interest to us here.} (This restriction is lifted in Sec.~\ref{sec:configurational}.) With this in mind, we can eliminate $\bm{v}$ from \eqref{eq:lagrangian_density_4}:
\begin{equation}
  \mathcal{L}
  = \rho\partial_t\Psi
    - \frac{1}{2}\rho \lvert \nabla\Psi \rvert^2
    - \rho\,e.
  \label{eq:lagrangian_density_5}
\end{equation}

On the other hand, if we vary $\mathcal{L}$ from  \eqref{eq:lagrangian_density_5} with respect to the density field,  we obtain
\begin{equation}
    \delta_{\rho}\mathcal{L}
    =
    \left(
    \partial_t\Psi
    -\frac{1}{2}|\nabla\Psi|^2
    -e
    -\rho\frac{\mathrm{d}e}{\mathrm{d}\rho}\right) \delta\rho.
\end{equation}
Requiring the variation $\delta_{\rho}\mathcal{L}$ vanish for arbitrary $\delta\rho$ leads to 
\begin{equation}
    \partial_t\Psi-\frac{1}{2}|\nabla\Psi|^2
    =
    e+\rho\frac{\mathrm{d}e}{\mathrm{d}\rho}.
    \label{eq:potential_interms_of_e}
\end{equation}
Interestingly, this relationship is a form of what Sedov \cite[p.~156]{Sedov1972AMechanics} terms the ``Cauchy--Lagrange integral'' (see discussion in \cite{Christov2007ModelingPropagation} and \eqref{cauchy-lagrange integral} below). To the best of our knowledge, \eqref{eq:potential_interms_of_e} is not derived in this way in the literature; instead, it is usually obtained as a first integral of the momentum equation derived from the Euler--Lagrange equations~\eqref{eq:EL operator}.

With this in mind, we may use \eqref{eq:potential_interms_of_e} to replace $\partial_t\Psi - \tfrac{1}{2}|\nabla\Psi|^2$ in \eqref{eq:lagrangian_density_5}, whereupon the internal energy cancels:
\begin{equation}
  \mathcal{L}
  = \rho^2 \frac{\mathrm{d}e}{\mathrm{d}\rho}.
  \label{eq:lagrangian_density_6}
\end{equation}
Equation~\eqref{eq:lagrangian_density_6} is as far as we can go at this stage. To make further progress. 

To begin, let us recall some thermodynamic relations (see, e.g., \cite{Blackstock2000FundamentalsAcoustics,Pierce2019Acoustics,Gonzalez2001AMechanics} and, instructively, \cite{Seliger1968VariationalMechanics}), and then specialize to perfect gases. For an \emph{isentropic process} ($\mathrm{d}s=0$), the fundamental thermodynamic identities relating the specific internal energy $e$, the specific enthalpy $h \equiv e + p/\rho$, the density $\rho$, and the pressure $p$ are:
\begin{subequations}\label{eq:d_thermo}\begin{alignat}{3}
    \mathrm{d}e &= - p \, \mathrm{d}(1/\rho) &\qquad &\Rightarrow &\qquad p &= \rho^2 \frac{\mathrm{d} e}{\mathrm{d} \rho}, \label{eq:d_thermo_de}\\
    \mathrm{d}h &= (1/\rho) \mathrm{d}p &\qquad &\Rightarrow &\qquad  \frac{\mathrm{d} h}{\mathrm{d} p} &= \frac{1}{\rho}. \label{eq:d_thermo_dh}
\end{alignat}\end{subequations}
The homentropic assumption makes each thermodynamic variable a function of any other, so the derivatives above are written as total ones throughout this section; a subscript, as in $|_{\rho=\rho_0}$, indicates evaluation at the ambient state, and at a general state the sound speed is $c^2=\mathrm{d}p/\mathrm{d}\rho$. For a perfect gas, which obeys $p=\rho R T$ and $\mathrm{d}e = c_v\,\mathrm{d}T$, a lengthy calculation using \eqref{eq:d_thermo_de} yields the \emph{isentropic (polytropic) relations}
\begin{equation} 
    p = p_0\left(\frac{\rho}{\rho_0}\right)^\gamma 
    \quad \Rightarrow \quad 
    c^2 = \frac{\mathrm{d} p}{\mathrm{d} \rho} = \frac{\gamma p}{\rho} 
    \quad \Rightarrow \quad 
    \rho = \rho_0 \left(\frac{c^2}{c_0^2}\right)^{\frac{1}{\gamma-1}}, 
\label{eq:isentropic_rels} 
\end{equation} 
where $\gamma = c_p/c_v$ is the ratio of specific heats. 

Now, substituting \eqref{eq:d_thermo_de} into \eqref{eq:lagrangian_density_6}, we obtain
\begin{equation}
    \mathcal{L}=p.
    \label{eq:Lagrangian_is_the_pressure}
\end{equation}
Thus, the Lagrangian density may be written entirely in terms of the pressure. Indeed, Seliger and Whitham \cite{Seliger1968VariationalMechanics} famously remark that ``\textit{the Lagrangian density is just the pressure!}'' (emphasis theirs); see also \cite{Tong2025FluidMechanics}.

Finally, while \eqref{eq:Lagrangian_is_the_pressure} holds for any homentropic barotropic fluid, for the case of a perfect gas, we use the polytropic relations \eqref{eq:isentropic_rels} to rewrite the Lagrangian density solely in terms of the sound speed:
\begin{equation}
    \mathcal{L}
    =
    \frac{\rho_0}{\gamma(c_0^2)^{\frac{1}{\gamma-1}}}
    (c^2)^{\frac{\gamma}{\gamma-1}}.
    \label{eq:final lagrangian density}
\end{equation}
This is precisely the Lagrangian density posited in \cite{Christov2015CorrigendumPropagation}, and we have supplied its derivation starting from the physical Lagrangian \eqref{eq:lagrangian_density}.

\subsection{Wave equations emerging from the Eulerian variational principle}
\label{sec:perturbed_lagrangians}

Given the underlying physical significance of the variational principle, and the desire to understand what Morse and Ingard were after (recall Sec.~\ref{sec:Eulerian_background}), as well as the origin of the corrected Lagrangian density $\mathcal{L}_\mathrm{corr}$ from \eqref{eq:christov_jordan_corrected_lagrangian}, we take the point of view of perturbation theory applied to the variational principle. This approach ensures that any approximation made still yields governing equations with a variational structure. Regarding our bookkeeping throughout: since the Euler--Lagrange operator differentiates once, a Lagrangian density retained through order $n+1$ yields equations accurate to order $n$. 

\subsubsection{Perturbation expansion of the Lagrangian density}
\label{sec:peturbation_Eulerian}

To this end, we rewrite \eqref{eq:final lagrangian density} as 
\begin{equation}
    \mathcal{L}=K\,(c^{2})^{m},
    \qquad
    m = \frac{\gamma}{\gamma-1},
    \qquad
    K = \frac{\rho_0}{\gamma(c_0^2)^{\frac{1}{\gamma-1}}},
    \label{eq:lagrangian_density_for_expansion}
\end{equation}
where $K$ and $m$ are constants independent of the field $\Psi$. To proceed, we need to furnish a relation between $c^2$ and $\Psi$. Returning to \eqref{eq:potential_interms_of_e}, and recalling \eqref{eq:d_thermo_de} and the definition of enthalpy, it is obvious that 
\begin{equation}
    \partial_t\Psi-\frac{1}{2}|\nabla\Psi|^2
    =
    h.
    \label{relation 0}
\end{equation}
Then, using \eqref{eq:d_thermo_dh} and \eqref{eq:isentropic_rels}$_1$, $h =\int_0^h \mathrm{d}\tilde{h} = \int_{\rho_0}^\rho \mathrm{d}p(\tilde{\rho})/\tilde{\rho} = (c^2 - c_0^2)/(\gamma-1)$, having chosen the reference enthalpy $h_0=0$ without loss of generality (also derived in \cite[p.~86]{Blackstock2000FundamentalsAcoustics} but with the opposite sign convention on the potential; and in \cite{Christov2007ModelingPropagation}). Finally, using \eqref{relation 0} and the relationship between $h$ and $c^2$, \eqref{eq:lagrangian_density_for_expansion} becomes 
\begin{equation}
    \begin{aligned}
    \mathcal{L}
    &=
    K\left[c_{0}^{2}+(\gamma-1)\left(\partial_{t}\Psi-\frac{1}{2}|\nabla\Psi|^{2}\right)\right]^{m}\\
    &=
    K(c_{0}^{2})^{m}
    \left[
    1 + \underline{\frac{\gamma-1}{c_{0}^{2}} \left(\partial_{t}\Psi-\frac{1}{2}|\nabla\Psi|^{2}\right)}
    \right]^{m}.
    \end{aligned}
\label{eq:L_before_binomial}
\end{equation}
Equation~\eqref{eq:L_before_binomial} is exact for a perfect gas: substituted into \eqref{eq:EL operator}, it yields a ``fully nonlinear'' acoustic wave equation in terms of $\Psi$ \cite{Christov2015CorrigendumPropagation,Rasmussen2008AnalyticalDissipation}. Our interest here, however, is in the weakly nonlinear models obtained by truncation, to which we now turn.

A ``fully nonlinear'' acoustic wave equation, in terms of the potential $\Psi$, emerges from substituting \eqref{eq:L_before_binomial} into the Euler--Lagrange equation~\eqref{eq:EL operator} (see also \cite{Christov2015CorrigendumPropagation,Rasmussen2008AnalyticalDissipation}). This equation can be analyzed to understand nonlinear acoustic wave propagation. 

However, there has been a general trend and desire in the acoustics literature to obtain so-called ``weakly nonlinear'' governing equations (termed ``model'' equations of nonlinear acoustics) through a zoo of approximations \cite{Crighton1979ModelAcoustics,Makarov1997NonlinearII,Blackstock1997SomeAcoustics,Hamilton2024ModelEquations,Jordan2016A19102009}. To this end, a binomial expansion under the assumption that the underlined term in \eqref{eq:L_before_binomial} is small compared to $1$, yields
\begin{multline}
    \mathcal{L}
    =
    \underbrace{K(c_{0}^{2})^{m}}_{=\rho_0 c_0^2/\gamma}
    \Bigg[
    1
    +\frac{\gamma}{c_{0}^{2}}
    \left(
    \partial_{t}\Psi-\frac{1}{2}|\nabla\Psi|^{2}
    \right)
    +\frac{\gamma}{2c_{0}^{4}}
    \left(
    \partial_{t}\Psi-\frac{1}{2}|\nabla\Psi|^{2}
    \right)^{2}\\
    +\frac{\gamma(2-\gamma)}{6c_{0}^{6}}
    \left(
    \partial_{t}\Psi-\frac{1}{2}|\nabla\Psi|^{2}
    \right)^{3}
    + \cdots 
    \Bigg].
\label{eq:L_substituted_Q}
\end{multline}

Although we obtained \eqref{eq:L_substituted_Q} via the perfect-gas relations, no equation of state is actually required. Equations \eqref{eq:Lagrangian_is_the_pressure} and \eqref{relation 0} together say that $\mathcal{L}=p(h)$ with $h=\partial_t\Psi-\tfrac12|\nabla\Psi|^2$, so the expansion above is nothing more than a Taylor series of the pressure in the specific enthalpy. Using \eqref{eq:d_thermo_dh} repeatedly, $\mathrm{d}p/\mathrm{d}h=\rho$, $\mathrm{d}^2p/\mathrm{d}h^2=\rho/c^2$, and $\mathrm{d}^3p/\mathrm{d}h^3=(\rho/c^4)\big[1-(\rho/c^2)\mathrm{d}^2p/\mathrm{d}\rho^2\big]$. As is common in nonlinear acoustics \cite{Hamilton2024ModelEquations}, we introduce the \emph{nonlinearity parameter} $B/A$ of the fluid \cite{Beyer2024TheB/A}, defined as a  thermodynamic derivative:
\begin{equation}
    \frac{B}{A}
    =
    \frac{\rho_0}{c_0^{2}}
    \left.\frac{\mathrm{d}^{2} p}{\mathrm{d}\rho^{2}}\right|_{\rho=\rho_0}.
    \label{eq:BoverA}
\end{equation}
Then, the expansion of \eqref{eq:Lagrangian_is_the_pressure} about the ambient state reads
\begin{multline}
    \mathcal{L}
    =
    p_0
    +\rho_0 \left( \partial_t\Psi-\frac{1}{2}|\nabla\Psi|^{2} \right)
    +\frac{\rho_0}{2c_0^{2}} \left( \partial_t\Psi-\frac{1}{2}|\nabla\Psi|^{2} \right)^{2}\\
    +\frac{\rho_0}{6c_0^{4}}\left(1-\frac{B}{A}\right) \left( \partial_t\Psi-\frac{1}{2}|\nabla\Psi|^{2} \right)^{3}
    +\cdots,
    \label{eq:L_general_expansion}
\end{multline}
for any nondissipative homentropic flow. Equation~\eqref{eq:L_substituted_Q} is the perfect-gas version, for which $B/A=\gamma-1$. As a point of reference, air has $B/A\approx0.4$ (consistent with
$\gamma=1.4$), whereas distilled water at $20^\circ$C has $B/A\approx5.0$ and
ethanol $\approx10.5$ \cite[Sec.~6]{Beyer2024TheB/A}, well outside the range a
perfect-gas closure can reach; this is why we have kept the expansion general.

\subsubsection{Linear acoustics}

To recover the linear theory, we retain only terms up to \emph{quadratic} order in the field amplitude (and its derivatives). Since $|\nabla\Psi|^{2}$ is already quadratic, we have
\begin{equation}
    \left(
    \partial_{t}\Psi-\frac{1}{2}|\nabla\Psi|^{2}
    \right)^{2}
    =
    (\partial_{t}\Psi)^{2}
    +\text{cubic terms}.
\end{equation}
Hence, \eqref{eq:L_general_expansion} becomes
\begin{equation}
    \mathcal{L} = p_0 + \rho_0\,\partial_t\Psi
    - \frac{\rho_0}{2}|\nabla\Psi|^2
    + \frac{\rho_0}{2c_0^2}(\partial_t\Psi)^2
    + \text{cubic terms}.
    \label{eq:L_quad_raw}
\end{equation}

The constant term does not contribute to the Euler--Lagrange equation, while the $\partial_t\Psi$ term is a time derivative and may also be discarded, as discussed above. Therefore, up to the irrelevant overall multiplicative constant $\rho_0$, the Lagrangian density to quadratic order reduces to
\begin{equation}
    \mathcal{L}_2
    =
    \frac{1}{2c_{0}^{2}}(\partial_{t}\Psi)^{2}
    -\frac{1}{2}|\nabla\Psi|^{2}.
\label{eq:L_linearized_from_final}
\end{equation}
Equation~\eqref{eq:L_linearized_from_final} is the Lagrangian density from \eqref{eq:morse_ingard_lagrangian_component}, up to rescaling. Applying the Euler--Lagrange operator \eqref{eq:EL operator} with $\mathcal{L}=\mathcal{L}_2$ yields the linear acoustic wave equation~\eqref{eq:morse_ingard_wave_equation}. Thus, the standard linear acoustic wave equation can be recovered by a perturbative expansion of the ``fully nonlinear'' Lagrangian density \eqref{eq:Lagrangian_is_the_pressure}.

\subsubsection{Weakly nonlinear acoustics}

We now keep higher-order terms in the expansion of the Lagrangian density. Working from the general form \eqref{eq:L_general_expansion}, so that no equation of state is assumed, and discarding constant terms and the total time derivative, the Lagrangian density to \emph{cubic} order reduces, up to the irrelevant overall factor $\rho_0$, to
\begin{equation}
    \mathcal{L}_3
    =
    \frac{1}{2c_0^2}(\partial_t\Psi)^2
    -\frac{1}{2}|\nabla\Psi|^2
    -\frac{1}{2c_0^2}(\partial_t\Psi) |\nabla\Psi|^2
    +\frac{1}{6c_{0}^{4}}\left(1-\frac{B}{A}\right)(\partial_t\Psi) ^{3},
    \label{eq:cubic_truncated_lagrangian}
\end{equation}
which is the same truncated Lagrangian density discussed in \cite[Eq.~(11)]{Rasmussen2008AnalyticalDissipation} up to a multiplicative factor. By starting from the basic variational principle, based on the physical Lagrangian density \eqref{eq:lagrangian_density}, manipulating it into the exact Lagrangian density \eqref{eq:final lagrangian density} in terms of $\Psi$, then performing the perturbation expansion and truncation, we have explained the origin of  \eqref{eq:christov_jordan_corrected_lagrangian}, as we set out to do. 

It is pedagogical to compare \eqref{eq:cubic_truncated_lagrangian} with the guessed $\mathcal{L}_{\mathrm{MI}}$ of \eqref{eq:morse_ingard_nonlinear_lagrangian}. Expanding the latter and dividing by $\rho_0$, the quadratic terms and the cross term $-(\partial_t\Psi)|\nabla\Psi|^2/2c_0^2$ agree exactly. The sole discrepancy is the coefficient of $(\partial_t\Psi)^3$: $\mathcal{L}_{\mathrm{MI}}$ of \eqref{eq:morse_ingard_nonlinear_lagrangian} forces $1/2c_0^4$, whereas $\mathcal{L}_3$ from \eqref{eq:cubic_truncated_lagrangian} gives $(1-B/A)/6c_0^4$. The ratio of the two is $\tfrac{1}{3}(1-B/A)$, which for a perfect gas is the factor $\tfrac{1}{3}(2-\gamma)$ that Christov and Jordan \cite{Christov2015CorrectionsGases} identified as missing from Morse and Ingard's coefficient. Writing it in terms of $B/A$ shows that the two coefficients agree only if $B/A=-2$, i.e., if the \emph{coefficient of nonlinearity} $1+B/(2A)$ \cite{Beyer2024TheB/A} of the fluid vanishes! In other words, $\mathcal{L}_{\mathrm{MI}}$ is the correct Lagrangian density for a fictitious fluid of vanishing nonlinearity. Indeed, this coefficient would have been hard to guess without specifying the equation of state and performing the weakly nonlinear expansion.

Now, applying the Euler--Lagrange operator \eqref{eq:EL operator} with $\mathcal{L} = \mathcal{L}_3$, we obtain the following weakly nonlinear acoustic wave equation:
\begin{equation}
    \left[
    1+\frac{1}{c_0^2}\left(1-\frac{B}{A}\right)\partial_t\Psi
    \right]
    \partial_t^2\Psi
    -
    \left(
    c_0^2+\partial_t\Psi
    \right)\nabla^2\Psi
    =
    2\nabla\Psi\cdot\nabla(\partial_t\Psi),
    \label{eq: weakly nonlinear equation from L3}
\end{equation} 
which is the nondissipative version of the equation of Rasmussen et al.~\cite[Eq.~(7)]{Rasmussen2008AnalyticalDissipation} (or \cite[Eq.~(1)]{Rasmussen2010AnalyticalDissipation}), likewise expressed there in terms of $B/A$. For a perfect gas, $1-B/A = 2-\gamma$, so \eqref{eq:cubic_truncated_lagrangian} reduces to the Lagrangian density \eqref{eq:christov_jordan_corrected_lagrangian} of Christov and Jordan \cite{Christov2015CorrectionsGases}, up to the overall factor $\rho_0$.

We conclude our discussion of weakly nonlinear acoustics by connecting with the recent work by Scholle et al.~\cite{Scholle2023AEquations}, who presented a variational perturbation framework in which the Lagrangian density is expanded formally using a generalized Taylor series to successive orders in chosen perturbation variables. Substituting \eqref{eq:L_before_binomial} into the weakly nonlinear perturbation Lagrangian $\Omega_{\mathrm{wnl}}$ proposed in \cite{Scholle2023AEquations}, one recovers, up to an irrelevant overall constant factor, the same cubic truncated Lagrangian density given in \eqref{eq:cubic_truncated_lagrangian}. But we assert that our approach in this chapter is more pedagogically transparent. 

Another point worth noting concerns the Lagrangian density for irrotational barotropic flow posited in \cite[Eq.~(9)]{Scholle2023AEquations}, described therein as a ``simplified form'' of the variational principle introduced by Seliger and Whitham \cite{Seliger1968VariationalMechanics}. However, it is not obvious how this simplification occurred. It would appear that the simplest Lagrangian density, $\mathcal{L} = p$ from \eqref{eq:Lagrangian_is_the_pressure}, was rewritten as $\mathcal{L} = \rho h - \rho e$ and then \eqref{relation 0} was used to eliminate $h$ from the latter. Furthermore, \cite[Eq.~(9)]{Scholle2023AEquations} appears to be identical (up to sign convention on $\Psi$) to the Lagrangian density posited in \cite[p.~304]{Fetter2003TheoreticalContinua}, which was introduced without proof by way of saying that ``it is remarkable that'' it produces the expected continuity and momentum equations.

\section{Variational principle for acoustics in the Lagrangian description}
\label{sec:configurational}

The discussion in Sec.~\ref{sec:Eulerian} was developed entirely from the Eulerian description of continuum mechanics. In this section, we adopt the Lagrangian viewpoint and examine a different symmetry of the action, namely material relabeling symmetry, which was not considered by Eckart~\cite{Eckart1960VariationHydrodynamics}, and leads to a pseudomomentum balance. Notable work on relabeling symmetry can be found in \cite{Salmon1988HamiltonianMechanics, Badin2018VariationalDynamics, Padhye1996FluidSymmetry, Vallis2006AtmosphericDynamics}. Importantly, the Lagrangian viewpoint obviates the need for \textit{ad hoc} fixes and extra constraints related to conservation of mass or the rotational part of the velocity field, meaning that, until stated otherwise, flows in this section are \emph{not} assumed irrotational. But we do assume a flat Euclidean space. Furthermore, the thermodynamic assumptions of Sec.~\ref{sec:Eulerian_background} remain in force throughout this section: the fluid is nondissipative and homentropic, hence it obeys a barotropic equation of state. Since the Lagrangian density \eqref{eq:compressible lagrangian} below is stated for a general $e(\rho,s)$, we write partial derivatives at fixed $s$ in this section, in place of the total derivatives of Sec.~\ref{sec:Eulerian}; for an evaluation point such as $|_{\rho=\bar\rho}$, fixed $s$ is left understood.

Before we begin, it is worth clarifying what our contributions are and what we take directly from the literature. The geometric setting, the variational identities relating the two variation operators, and the decomposition of $\delta A$ into a temporal charge, a spatial flux, and a bulk Euler--Lagrange term are all due to Singh and Hanna \cite{Singh2021Pseudomomentum:Consequences}, who developed them for a general material Lagrangian density and applied them to elastic solids, incompressible ideal fluids, and thin structures. Our contribution is the compressible, homentropic specialization. While in \cite{Singh2021Pseudomomentum:Consequences}, the pressure is a multiplier enforcing $J=1$, here mass conservation holds identically through $\rho=\bar\rho J^{-1}$, and the pressure is not an independent field at all, entering only through the derivatives of $e$. What this yields that is new is the boundary and jump conditions accompanying the three balances and the perturbative treatment of Sec.~\ref{sec:Lagrangian_weakly_nonlinear}. The bulk balances for a compressible fluid go back to Berdichevsky \cite{Berdichevsky2009VariationalFundamentals}, who obtained the pseudomomentum balance (though not under that name) by projecting the momentum balance onto the deformation gradient.

\subsection{Background and useful preliminaries}
\label{background and useful preliminaries}

\begin{figure}[t]
    \centering
    \includegraphics[width=\textwidth]{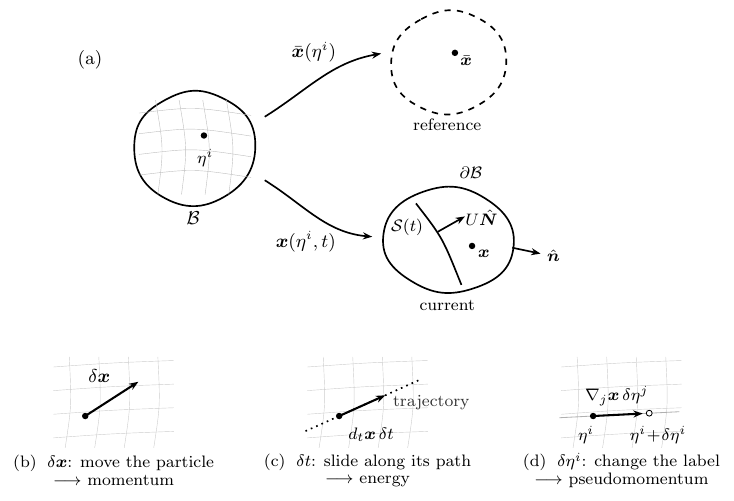}
    \caption{The configurational setting. (a) The body $\mathcal{B}$, its material labels $\eta^i$, and the two maps carrying them to the reference and current configurations. (b)--(d) The three independent variations, each shown at a single particle against the convected material grid: in (c) the dotted curve is the particle's trajectory, and in (d) the darker line is the material coordinate line through it, along which the open circle marks the neighboring particle $\eta^i+\delta\eta^i$.}
    \label{fig:configurational_setting}
\end{figure}

Throughout this section, we employ the notation and machinery introduced in \cite{Singh2021Pseudomomentum:Consequences}. To briefly summarize---consider a body $\mathcal{B}$ to be a differentiable, orientable manifold with boundary $\partial\mathcal{B}$. Material elements of $\mathcal{B}$ are labeled by material coordinates $\eta^i$. A configuration of the body is described by a position field $\bm{x}(\eta^i,t)\in\mathbb{E}^3$ in Euclidean space, while the reference configuration is represented by the time-independent map $\bar{\bm{x}}(\eta^i)$; see Fig.~\ref{fig:configurational_setting}(a). The particle velocity is then $d_t\bm{x}$. For simplicity,  we write $d_t$ for the material time derivative, and we write $\nabla_i$ and $\bar\nabla_i$ for the covariant derivatives constructed from the present and reference metrics, whose components in the present and reference bases are $g_{ij}=\nabla_i\bm{x}\cdot\nabla_j\bm{x}$  and $\bar g_{ij}=\bar\nabla_i\bar{\bm{x}}\cdot\bar\nabla_j\bar{\bm{x}}$, respectively; since  $\bm{x}$ is written in vector form (and not in index notation) on $\mathcal{B}$, $\nabla_i\bm{x}=\partial\bm{x}/\partial \eta^i$ carries no connection terms, and likewise for $\bar\nabla_i\bar{\bm{x}}$. Two identities follow at once and will be used repeatedly. First, since $\nabla^i\bm{x}=g^{ik}\nabla_k\bm{x}$, contraction with the definition of the present metric gives
\begin{equation}
    \nabla^{i}\bm{x}\cdot\nabla_j\bm{x}=\delta^{i}_{j}.
    \label{eq:tangent_identity}
\end{equation}
Second, for a \emph{space-filling} body (i.e., one whose material manifold is three-dimensional and embedded in flat $\mathbb{E}^3$), the tangent vectors $\nabla_i\bm{x}$ span all of $\mathbb{E}^3$, so the normal component vanishes identically and hence
\begin{equation}
    g^{ij}\nabla_i\nabla_j\bm{x} = \nabla_i\nabla^i\bm{x}=
    \nabla^{2}\bm{x}=\bm{0}.
    \label{eq:flatness_identity}
\end{equation}
A third identity we shall need is that for the derivative of a determinant \cite[Sec.~2.4.1]{Gonzalez2001AMechanics}: for any operator $D$ acting at fixed material label, $DJ = J\,\nabla^i\bm{x}\cdot\nabla_i(D\bm{x})$; we use it below with $D=\tilde\delta$ and with $D=d_t$. None of these identities involves an approximation or a particular choice of material labels. We consider a fluid body to be space-filling by construction. We emphasize that the present subsection is purely kinematic and geometric: the identities \eqref{eq:tangent_identity} and  \eqref{eq:flatness_identity} involve no thermodynamic or constitutive assumption and hold for any Lagrangian density.

Let the Lagrangian density in the reference configuration be $\bar{\mathcal{L}}(\eta^i,t;\bm{x},\nabla_i\bm{x},d_t\bm{x})$ and let the corresponding quantity in the current configuration be $\mathcal{L}(\eta^i,t;\bm{x},\nabla_i\bm{x},d_t\bm{x})=J^{-1}\bar{\mathcal{L}}$, the two being related by the ratio of volume elements. The arguments preceding the semicolon are regarded as independent variables, while those following it are treated as dependent fields. The dependence on $\nabla_i\bm{x}$ enters through the metric, hence through $J=\sqrt{g/\bar g}$ and, by conservation of mass, through the density $\rho$. The action written over the reference configuration then takes the form
\begin{equation}
    A 
    = \int_{t_0}^{t_1} \mathrm{d}t \int_{\mathcal{B}} \mathrm{d}\bar V \,
    \bar{\mathcal{L}} 
    = \int_{t_0}^{t_1} \mathrm{d}t \int_{\mathcal{B}} \mathrm{d}V \,
    J^{-1} \bar{\mathcal{L}},
    \label{eq:action}
\end{equation}
where $J$ is the Jacobian relating the present and reference volume elements.

We assume the manifold carries the metric, so that the present and reference volume elements are $\mathrm{d}V=\sqrt{g}\,\mathrm{d}\eta^1\,\mathrm{d}\eta^2\,\mathrm{d}\eta^3$ and $\mathrm{d}\bar V=\sqrt{\bar g}\,\mathrm{d}\eta^1\,\mathrm{d}\eta^2\,\mathrm{d}\eta^3$, respectively. Here $g=\det(g_{ij})$ and $\bar g=\det(\bar g_{ij})$, and we use the standard relation $\mathrm{d}\bar V=J^{-1}\mathrm{d} V$. We now investigate the transformation of the action~\eqref{eq:action} under the infinitesimal change of variables
\begin{subequations}\label{eq:transformation}\begin{align}
    \eta'^i &= \eta^i+\delta\eta^i(\eta^j,t),\\
    t' &= t+\delta t,\\
    \bm{x}'(\eta'^i,t') &= \bm{x}(\eta^i,t)+\delta\bm{x}(\eta^i,t).
\end{align}\end{subequations}

Before proceeding with the calculation, we first state a few standard results that will be used in what follows. Consider a Lagrangian density $\mathcal{L}(\phi_\alpha,\phi_{\alpha,\nu},z^{\mu})$ depending on field(s) $\phi_\alpha$ and their derivatives $\phi_{\alpha,\nu}=\partial\phi_\alpha/\partial z^{\nu}$ over a spacetime domain $\Omega$ with coordinates $z^{\mu}$ (treated as generic placeholders, distinct from the material label $\eta^i$ and position $\bm{x}$ used elsewhere). Under an infinitesimal perturbation of the spacetime integration domain $\Omega\mapsto\Omega'$, the corresponding change in the action~\cite{Goldstein2002ClassicalMechanics} is
\begin{multline}
    \int_{\Omega'} \mathcal{L} \left(\phi'_\alpha,\phi'_{\alpha,\nu},z'^{\mu}\right)\,\mathrm{d}^4z'
    -\int_{\Omega} \mathcal{L} \left(\phi_\alpha,\phi_{\alpha,\nu},z^{\mu}\right)\,\mathrm{d}^4z
    \\
    =
    \int_{\Omega} \frac{\partial}{\partial z^{\nu}}
    \left[
    \frac{\partial \mathcal{L}}{\partial \phi_{\alpha,\nu}}\,\tilde{\delta}\phi_\alpha
    + \mathcal{L} \left(\phi_\alpha,\phi_{\alpha,\nu},z^{\mu}\right)\,\delta z^{\nu}
    \right]\mathrm{d}^4z,
    \label{eq:standard result 1}
\end{multline}
with summation over $\alpha$ implied. 
Here, we have introduced a second variation symbol, $\tilde{\delta}$. The distinction between $\tilde{\delta}$ and $\delta$ is central to what follows. Writing $\delta$ for the total change in a quantity when the label, the time, \emph{and} the position are all perturbed by \eqref{eq:transformation}, $\tilde{\delta}$ denotes the change measured while following a fixed material element, that is, at fixed $\eta^i$ and $t$. It is $\tilde{\delta}$ that isolates the motion of the body from a mere relabeling of its points, and \eqref{eq:standard_result_3} below is the bookkeeping that relates the two. Because $\tilde{\delta}$ holds the label fixed, it commutes with derivatives taken at fixed label, which gives another standard result:
\begin{equation}
    \tilde{\delta} \left(\frac{\mathrm{d}\mathcal{L}}{\mathrm{d}z^{\mu}}\right)
    =
    \frac{\mathrm{d}}{\mathrm{d}z^{\mu}} \left(\tilde{\delta}\mathcal{L}\right).
    \label{eq:standard_result_2}
\end{equation}

We now specialize the generic identities \eqref{eq:standard result 1} and \eqref{eq:standard_result_2} to the fluid problem, taking the field to be the position, $\phi_\alpha\to\bm{x}$, and the generic coordinates to be the material label and time, $z^\mu\to(\eta^i,t)$, so that $\mathrm{d}^4z\to\mathrm{d}t\,\mathrm{d}V$ and the generic divergence $\partial/\partial z^\nu$ resolves into a temporal part $d_t$ and a spatial part $\nabla_i$. Using the chain rule and \eqref{eq:standard_result_2}, it is then straightforward to show that $\delta\bm{x}(\eta^i,t)$ and $\tilde{\delta}\bm{x}(\eta^i,t)$ are related by
\begin{equation}
    \tilde{\delta}\bm{x}(\eta^i,t)
    =
    \delta\bm{x}(\eta^i,t)
    -\nabla_j\bm{x}(\eta^i,t)\,\delta\eta^j
    -d_t\bm{x}(\eta^i,t)\,\delta t,
    \label{eq:standard_result_3}
\end{equation}
which is also reported in \cite{Singh2021Pseudomomentum:Consequences} and is illustrated in Fig.~\ref{fig:configurational_setting}(b)--(d), where each of the three variations is shown acting on a single particle. Now, using \eqref{eq:standard result 1}, \eqref{eq:standard_result_2}, and the fact that $J^{-1}\mathrm{d}V$ is independent of both $t$ and the variation, we can show that the variation of the action~\eqref{eq:action} under the infinitesimal change of variables \eqref{eq:transformation} is
\begin{equation}
    \delta A
    =\int_{t_0}^{t_1}  \mathrm{d}t \int_{\mathcal{B}}   \mathrm{d}V \,
    \Big[
    J^{-1} d_t \big(\bar{\mathcal{L}}\,\delta t\big)
    +\nabla_i \big(J^{-1}\bar{\mathcal{L}}\,\delta\eta^i\big)
    +J^{-1}\tilde{\delta}\bar{\mathcal{L}}
    \Big].
\label{eq:var_action_standard_format}
\end{equation}

Recall also the generalized divergence theorem and the generalized transport theorem (Leibniz rule) applied to an arbitrary vector field $\bm{A}$:
\begin{align}
    \int_{\mathcal{B}} \mathrm{d}V\, \nabla\cdot\bm{A}
    &=
    \int_{\partial\mathcal{B}} \mathrm{d}A\, \hat{\bm{n}}\cdot\bm{A}
    -\int_{\mathcal{S}(t)} \mathrm{d}A\, \llbracket \hat{\bm{N}}\cdot\bm{A} \rrbracket,
    \label{eq:gen_div_theorem}\\
    d_t \int_{\mathcal{B}} \mathrm{d}V\, J^{-1}\bm{A}
    &=
    \int_{\mathcal{B}} \mathrm{d}V\, J^{-1} d_t\bm{A}
    -\int_{\mathcal{S}(t)} \mathrm{d}A\, \llbracket J^{-1}U\bm{A}\rrbracket,
\label{eq:gen_transport_theorem}
\end{align}
allowing for jump discontinuities across a propagating interface $\mathcal{S}(t)$. Here, $\hat{\bm{N}}$ is the unit normal to the surface $\mathcal{S}(t)$, $\llbracket f\rrbracket = f^{+}-f^{-}$ denotes the jump across it, with $\hat{\bm{N}}$ directed from the $-$ side to the $+$ side, and $\hat{\bm{n}}$ the unit outward normal to the boundary $\partial\mathcal{B}$ of the volume $\mathcal{B}$. The scalar $U$ denotes the normal velocity of the propagating discontinuity (a ``weak shock front'' in acoustics).\footnote{Retaining such a discontinuity is consistent with the homentropic assumption made below because the entropy jump across a weak shock is third order in the shock strength \cite[\S86]{Landau1987FluidMechanics}, so the barotropic closure and the jump conditions derived from it remain accurate through second order.}

\subsection{Action, symmetries, and implied balance laws}
\label{action symmetries and implied balance}

As in \eqref{eq:lagrangian_density}, we take the Lagrangian density in the reference configuration for a compressible flow to be\footnote{To motivate \eqref{eq:compressible lagrangian}, recall $\mathcal{L}$ from \eqref{eq:lagrangian_density_4}. The last two terms combine into a total time derivative. Converting it to the referential Lagrangian as $\bar{\mathcal{L}}=J\mathcal{L}$, using \eqref{eq:material_com}, and the fact that $e$ is a scalar, we have $\bar{\mathcal{L}}=\tfrac{1}{2}\bar\rho\,d_t\bm{x}\cdot d_t\bm{x}-\bar\rho\,e(\rho,s)+\bar\rho\,d_t\Psi$. When variations are taken with respect to particle trajectories in the Lagrangian framework, $d_t\Psi$ contributes only a temporal boundary term that may be dropped.}
\begin{equation}
    \bar{\mathcal{L}}
    =\frac{1}{2}\bar{\rho} d_t\bm{x}\cdot d_t\bm{x}
    -\bar{\rho}\,e(\rho,s),
\label{eq:compressible lagrangian}
\end{equation}
where an overbar denotes a referential object. In the configurational context, there is no need to constrain the (Eulerian) fields by Lagrange multipliers, because here the variations are made in the trajectories of the fluid particles, as opposed to the Eulerian settings where it is simpler to take the variation of the fields independently (necessitating the constraints) \cite{Mobbs1982VariationalFlows, Salmon1988HamiltonianMechanics}. Here, $\bar{\rho}$ is the referential density of the fluid, which may be chosen to be uniform, and is related to the current density through the continuity constraint
\begin{equation}
    \rho = \bar{\rho}\,J^{-1}.
    \label{eq:material_com}
\end{equation}
Equation~\eqref{eq:material_com} is the material-frame statement of conservation of mass, equivalent to the Eulerian continuity equation enforced by the constraint in \eqref{eq:lagrangian_density}. Here it holds identically, by construction. (Its material rate form is given in \eqref{eq:continuity_material_frame} below.)
And, as before, $e(\rho,s)$ is the specific internal energy, and it has the same value in the reference and current configurations since it is a scalar. Note that \eqref{eq:compressible lagrangian} is stated for a general $e(\rho,s)$; the homentropic assumption is invoked only below, after \eqref{eq:com_act_variation}, where it removes the entropy variation and reduces $e(\rho,s)$ to $e(\rho)$. 

Given \eqref{eq:compressible lagrangian}, the action takes the form
\begin{equation}
    A
    =\int_{t_0}^{t_1} \mathrm{d}t \int_{\mathcal{B}}  \mathrm{d}V \, J^{-1}\bar{\mathcal{L}}
    =\int_{t_0}^{t_1} \mathrm{d}t \int_{\mathcal{B}}  \mathrm{d}V \,
    J^{-1}\left[\frac{1}{2}\bar{\rho}\, d_t\bm{x}\cdot d_t\bm{x}
    -\bar{\rho}\,e(\rho,s)\right].
\label{eq:comp_fluid_action}
\end{equation}
Taking its variation using \eqref{eq:var_action_standard_format}, we obtain
\begin{multline}
    \delta A
    =\int_{t_0}^{t_1}  \mathrm{d}t \int_{\mathcal{B}}   \mathrm{d}V \,
    \Bigg[
    J^{-1} d_t \Bigg(\Big[\tfrac{1}{2}\bar\rho d_t\bm{x}\cdot d_t\bm{x}
    -\bar\rho\,e(\rho,s)\Big]\delta t\Bigg)
    \\
    +\nabla_i \Bigg(J^{-1}\Big[\tfrac{1}{2}\bar\rho d_t\bm{x}\cdot d_t\bm{x}
    -\bar\rho\,e(\rho,s)\Big]\delta\eta^i\Bigg) \\
    +J^{-1}\Bigg(
    \bar\rho d_t\bm{x}\cdot d_t(\tilde{\delta}\bm{x})
    -\bar\rho\left.\frac{\partial e}{\partial\rho}\right|_s\tilde{\delta}\rho
    -\bar\rho\left.\frac{\partial e}{\partial s}\right|_{\rho}\tilde{\delta}s
    \Bigg)
    \Bigg].
\label{eq:com_act_variation}
\end{multline}
Observe that the pressure has not been introduced. It is neither an independent field nor a multiplier (like $\Psi$ in \eqref{eq:lagrangian_density}); $p$ appears below only through the derivatives of $e$. Whereas the Eulerian variational principle had introduced continuity via a multiplier, here \eqref{eq:material_com} holds by construction and the equation of state is the only additional ingredient.

For homentropic flow, the entropy is uniform and materially conserved, so $\tilde\delta s=0$ and the entropy term in \eqref{eq:com_act_variation} drops (note $(\partial e/\partial s)|_\rho=T\neq0$; it is the \emph{variation} that vanishes, not the derivative). The internal energy may then be written as $e(\rho)$, consistent with the barotropic closure introduced in Sec.~\ref{sec:Eulerian_background}. Then, applying $\tilde\delta$ to \eqref{eq:material_com}, the constrained variation of the density satisfies $\tilde{\delta}\rho=-\bar\rho J^{-1}\nabla^{i}\bm{x}\cdot \nabla_{i}\big(\tilde{\delta}\bm{x}\big)$. We also recall the thermodynamic identity $p=\rho^{2}(\partial e/\partial\rho)|_s$ from \eqref{eq:d_thermo_de}. Using these relations, and the product rule, \eqref{eq:com_act_variation} can be rewritten as
\begin{multline}
    \delta A
    =\int_{t_0}^{t_1}  \mathrm{d}t \int_{\mathcal{B}}   \mathrm{d}V \,
    \Bigg\{
    J^{-1} d_t \Big(\overbrace{
    \Big[\tfrac{1}{2}\bar\rho d_t\bm{x}\cdot d_t\bm{x}
    -\bar\rho\,e(\rho)\Big]\delta t
    +\bar\rho\, d_t\bm{x}\cdot\tilde{\delta}\bm{x}
    }^{Q}\Big) \\
    +\nabla_i \Big(\underbrace{
    J^{-1}\Big[\tfrac{1}{2}\bar\rho d_t\bm{x}\cdot d_t\bm{x}
    -\bar\rho\,e(\rho)\Big]\delta\eta^i
    +p\nabla^{i}\bm{x}\cdot\tilde{\delta}\bm{x}}_{\mathcal{J}^i}
    \Big)
    \\
    \underbrace{-\Big[J^{-1}d_t(\bar\rho d_t\bm{x})
    +\nabla_i(p \nabla^{i}\bm{x})\Big]}_{\bm{\mathcal{E}}} \cdot \tilde{\delta}\bm{x}
    \Bigg\},
\label{eq:schematic_deltaA}
\end{multline}
which fits the schematic form for $\delta A$ outlined by Singh and Hanna \cite{Singh2021Pseudomomentum:Consequences}. Here, as in \cite{Singh2021Pseudomomentum:Consequences}, $Q$ is the temporal boundary contribution, $\mathcal{J}^i$ is the associated spatial flux, and $\bm{\mathcal{E}}$ is the Euler--Lagrange (bulk) term multiplying the admissible variation $\tilde{\delta}\bm{x}$.

Using the generalized divergence \eqref{eq:gen_div_theorem} and transport \eqref{eq:gen_transport_theorem} theorems, and taking the variations to vanish at $t_0$ and $t_1$, we may rewrite \eqref{eq:schematic_deltaA} as
\begin{multline}
    \delta A
    =
    \int_{t_0}^{t_1} \mathrm{d}t \int_{\partial\mathcal{B}} \mathrm{d}A \,
    \hat{\bm{n}}\cdot\boldsymbol{\mathcal{J}}
    +\int_{t_0}^{t_1} \mathrm{d}t \int_{\mathcal{S}(t)}  \mathrm{d}A \,
    \llbracket -\hat{\bm{N}}\cdot\boldsymbol{\mathcal{J}}+J^{-1}UQ \rrbracket
    \\
    +\int_{t_0}^{t_1} \mathrm{d}t \int_{\mathcal{B}} \mathrm{d}V \,
    \boldsymbol{\mathcal{E}}\cdot\Big(\delta\bm{x}
    -\nabla_j\bm{x}\,\delta\eta^{j}
    -d_t\bm{x}\,\delta t\Big).
\label{eq:deltaA_boundary_shock_bulk}
\end{multline}
Now, consider the variation with respect to $\bm{x}$, meaning only $\delta \bm{x}$ is nonzero, and all other variations are zero.  Then, from the bulk and boundary terms, we obtain the following bulk and interfacial momentum balances, along with the boundary condition:
\begin{subequations}\begin{align}
    \boldsymbol{\mathcal{E}}(\bar{\mathcal{L}})
    =
    -\big[\nabla_i \big(p \nabla^{i}\bm{x}\big)
    + J^{-1} d_t \big(\bar\rho d_t\bm{x}\big)\big]
    &=\bm{0}
    \qquad \text{in } \mathcal{B},
    \label{eq: bulk momentum}\\
    \hat{n}_i\big(p \nabla^{i}\bm{x}\big) &= \bm{0}
    \qquad \text{on } \partial\mathcal{B},
    \label{eq:momentum bc} \\
    \left\llbracket -\hat{N}_i \big(p \nabla^{i}\bm{x}\big)
    +J^{-1}U\bar\rho d_t\bm{x} \right\rrbracket &= \bm{0}
    \qquad \text{on } \mathcal{S}(t).
    \label{eq:interfacial momentum balance}
\end{align}\end{subequations}
Equation~\eqref{eq: bulk momentum} is the bulk momentum balance, and \eqref{eq:interfacial momentum balance} is the interfacial momentum balance.  

Next, consider variations in time only, so that $\delta t\neq 0$ while all other variations, namely $\delta\bm{x}$ and $\delta\eta^i$, vanish. Collecting terms conjugate to $\delta t$ yields the bulk energy balance, the associated boundary condition, and the interfacial energy balance:
\begin{subequations}\begin{align}
    \boldsymbol{\mathcal{E}}(\bar{\mathcal{L}})\cdot d_t\bm{x}
    =
    -\Big[\nabla_i \big(p\nabla^{i}\bm{x}\big)
    + J^{-1} d_t \big(\bar\rho d_t\bm{x}\big)\Big]\cdot d_t\bm{x} &= 0
    \qquad \text{in } \mathcal{B},
    \label{eq:bulk_energy_balance}\\
    \hat{n}_i\big(p\nabla^{i}\bm{x}\big)\cdot d_t\bm{x} &= 0
    \qquad \text{on } \partial\mathcal{B},
    \label{eq:bc_energy_balance}\\
    \left\llbracket \hat{N}_i \big(p\nabla^{i}\bm{x}\big)\cdot d_t\bm{x}
    - U\Big(\tfrac{1}{2}\rho d_t\bm{x}\cdot d_t\bm{x}+\rho\,e(\rho)\Big)
    \right\rrbracket &= 0\qquad \text{on } \mathcal{S}(t).
    \label{eq:interface_energy_balance}
\end{align}\end{subequations}
Equation~\eqref{eq:bulk_energy_balance} is the bulk energy balance, obtained by projecting the bulk Euler--Lagrange operator onto the velocity $d_t\bm{x}$. Equation~\eqref{eq:interface_energy_balance} is the corresponding interfacial energy balance across $\mathcal{S}(t)$.

To cast the bulk energy balance~\eqref{eq:bulk_energy_balance} in a more recognizable form, let us expand the flux term by the product rule, i.e.,
\begin{equation}
    \nabla_i\big(p\nabla^{i}\bm{x}\big)\cdot d_t\bm{x}
    =
    \nabla_i\big(p\nabla^{i}\bm{x}\cdot d_t\bm{x}\big)
    -p\nabla^{i}\bm{x}\cdot\nabla_i\big(d_t\bm{x}\big).
    \label{eq:energy_flux_product_rule}
\end{equation}
Using the determinant identity of Sec.~\ref{background and useful preliminaries} with $D=d_t$, $d_t J = J\nabla^i\bm{x}\cdot\nabla_i(d_t\bm{x})$, which can be combined with the continuity constraint \eqref{eq:material_com} to yield a form of the material continuity equation 
\begin{equation}
    d_t \rho = -\rho \nabla^i \bm{x}\cdot \nabla_i(d_t \bm{x}).
    \label{eq:continuity_material_frame}
\end{equation}
Using \eqref{eq:continuity_material_frame} and the isentropic relation from \eqref{eq:d_thermo_de}, we convert the last term in \eqref{eq:energy_flux_product_rule} into
\begin{equation}
    -p \nabla^i\bm{x}\cdot\nabla_i(d_t\bm{x}) = p\frac{d_t\rho}{\rho}
    = \rho\frac{\partial e}{\partial\rho}\bigg|_s d_t\rho = \rho d_t e
    = J^{-1}d_t\big[\bar\rho\,e(\rho)\big],  
    \label{eq:internal_energy_rate}
\end{equation}
since $\bar\rho$ is time-independent. We may recast \eqref{eq:bulk_energy_balance} in the more recognizable form
\begin{equation}
    J^{-1} d_t \left[\tfrac{1}{2} \bar\rho  d_t\bm{x}\cdot d_t\bm{x}
    + \bar\rho\,e(\rho)\right]
    +\nabla_i \left(p\nabla^{i}\bm{x}\cdot d_t\bm{x}\right)=0
    \qquad \text{in } \mathcal{B},
\label{eq: energy balance final form}
\end{equation}
having further used $J^{-1}d_t(\bar\rho d_t\bm{x})\cdot d_t\bm{x}=J^{-1}d_t\big(\tfrac12\bar\rho d_t\bm{x}\cdot d_t\bm{x}\big)$ since $d_t\bar{\rho}=0$.

Finally, consider the variation of the material label coordinate, $\delta \eta^i \neq 0$, while all the other variations, namely $\delta \bm{x}$ and $\delta t$, vanish. We obtain the following pseudomomentum balance, boundary condition, and interfacial pseudomomentum balance:
\begin{subequations}\begin{align}
    \boldsymbol{\mathcal{E}}(\bar{\mathcal{L}})\cdot\nabla_j\bm{x}
    = -\big[\nabla_i(p\nabla^{i}\bm{x})\cdot\nabla_j\bm{x}
    + J^{-1}d_t\big(\bar\rho d_t\bm{x}\big) \cdot \nabla_j\bm{x}\big] &=0
    \qquad \text{in } \mathcal{B},
    \label{eq: bulk pseudomomentum balance}\\
    \hat{n}_i 
    \big[\tfrac{1}{2}\rho d_t\bm{x}\cdot d_t\bm{x}-\rho\,e(\rho) - p\big]\delta_j^i
     &= 0
    \qquad \text{on } \partial\mathcal{B},
    \label{eq:pseudomomentum_bc}\\
    \left\llbracket
    -\hat{N}_i \big[
    \tfrac{1}{2}\rho d_t\bm{x}\cdot d_t\bm{x}-\rho\,e(\rho) - p \big]\delta_j^i
    - U \rho d_t\bm{x}\cdot\nabla_j\bm{x}
    \right\rrbracket &= 0
    \qquad \text{on } \mathcal{S}(t),
    \label{eq:interface_pseudomomentum_balance}
\end{align}\end{subequations}
having employed \eqref{eq:tangent_identity}. Let us write $\bm{v}$ for the particle velocity $d_t\bm{x}$ and $v^2=\bm{v}\cdot\bm{v}$. Then, in passing, we note that \eqref{eq:pseudomomentum_bc} reveals the Eshelby, or energy--momentum, tensor for the compressible fluid, namely $(\tfrac{1}{2}\rho v^2 - \rho h ) \delta_j^i$, which is isotropic, since a nondissipative fluid cannot support shear stress. All three of \eqref{eq:momentum bc}, \eqref{eq:bc_energy_balance}, and \eqref{eq:pseudomomentum_bc} are \emph{natural} boundary conditions, holding only where the corresponding variation is left free on $\partial\mathcal{B}$: thus \eqref{eq:momentum bc} says that an unconstrained boundary is traction-free, and where the motion is instead prescribed (say, by a rigid wall), $\delta\bm{x}$ vanishes there and the condition is replaced by that essential one.

Expanding the flux term with \eqref{eq:tangent_identity} and \eqref{eq:flatness_identity} gives $\nabla_i(p\nabla^{i}\bm{x})\cdot\nabla_j\bm{x}=\nabla_j p$. Recalling that $\nabla_j$ carries no connection terms when acting on $\bm{x}$, it commutes with $d_t$. Hence, we have the identity
\begin{equation}
    d_t\big(d_t\bm{x}\cdot\nabla_j\bm{x}\big)
    =\big(d_t^2\bm{x}\big)\cdot\nabla_j\bm{x}
    +\nabla_j\big(\tfrac{1}{2}d_t\bm{x}\cdot d_t\bm{x}\big).
    \label{eq:dt_commute}
\end{equation}
Using the barotropic closure, which converts \eqref{eq:d_thermo_dh} into the material-gradient relation $\nabla_j p/\rho = \nabla_j h$ (in contrast with \eqref{eq:internal_energy_rate}, where a material \emph{rate} required only isentropy), recalling the specific enthalpy $h = e+p/\rho$, and employing \eqref{eq:dt_commute}, the balance~\eqref{eq: bulk pseudomomentum balance} may be divided through by $\rho$ and rewritten in the more recognizable form
\begin{equation}
    d_t \Big(d_t\bm{x}\cdot\nabla_{j}\bm{x}\Big)
    +\nabla_j \left[
    \frac{p}{\rho}
    -\tfrac{1}{2} d_t\bm{x}\cdot d_t\bm{x}
    +e(\rho)
    \right]
    = 0.
\label{eq: pseudomomentum recast}
\end{equation}

Equation~\eqref{eq: pseudomomentum recast} is also reported by Berdichevsky \cite[Eq.~(10.16), p.~459]{Berdichevsky2009VariationalFundamentals}. In that treatment, however, the internal energy density is allowed to depend explicitly on the spatial coordinates, and as a consequence, the pseudomomentum balance contains a source term. No such term appears in the present formulation, since here the internal energy density depends only on the constitutive variables and not explicitly on position. In the static limit, the pseudomomentum flux in \eqref{eq: pseudomomentum recast} is the specific enthalpy, which coincides with the configurational stress reported by Gurtin~\cite[Ch.~6(d)]{Gurtin2000ConfigurationalPhysics}.

\subsection{Consequences of the pseudomomentum balance}
\label{sec:consequences}

The foregoing bears on a longstanding debate regarding configurational balance laws. Maugin~\cite{Maugin2016ConfigurationalNumerics} took the view that pseudomomentum is the pull-back of momentum to the material reference configuration. Gurtin~\cite{Gurtin2000ConfigurationalPhysics}, in the context of evolving phase transitions and dissipative phenomena, invoked the second law of thermodynamics and argued, on the basis of material observer invariance, that configurational momentum obeys a balance \emph{independent} of the standard momentum balance. Rajagopal and Srinivasa~\cite{Rajagopal2005OnConfigurations}, on the other hand, regard the configurational momentum balance as a manifestation of evolving multiple natural configurations, requiring no additional law. 

The present calculation is instructive in this regard (Table~\ref{tab:balances}). The bulk momentum balance \eqref{eq: bulk momentum} is the statement $\bm{\mathcal{E}}(\bar{\mathcal{L}})=\bm{0}$, and the bulk energy \eqref{eq:bulk_energy_balance} and pseudomomentum \eqref{eq: bulk pseudomomentum balance} balances are its projections onto $d_t\bm{x}$ and $\nabla_j\bm{x}$; in the bulk they carry no independent information \cite{Singh2021Pseudomomentum:Consequences}. What the variational approach adds, in this source-free setting, is the accompanying boundary and jump conditions, which are genuinely distinct, together with the symmetry responsible for each.

\begin{table}[t]
\caption{The three balance laws implied by the action \eqref{eq:comp_fluid_action}, organized by the variation that generates each, with $\bm{v}=d_t\bm{x}$, $v^2=\bm{v}\cdot\bm{v}$, and $h=e+p/\rho$. The energy and pseudomomentum balances are projections of $\bm{\mathcal{E}}(\bar{\mathcal{L}})=\bm{0}$ in the bulk, but not on the boundary or across $\mathcal{S}(t)$.}
\label{tab:balances}
\begin{tabular}{l@{\quad}l@{\qquad}l@{\qquad}l}
\hline\noalign{\smallskip}
Variation & in $\mathcal{B}$ & on $\partial\mathcal{B}$ & across $\mathcal{S}(t)$ \\
\noalign{\smallskip}\svhline\noalign{\smallskip}
$\delta\bm{x}$
  & $\bm{\mathcal{E}}=\bm{0}$
  & $\hat{n}_i p \nabla^{i}\bm{x}=\bm{0}$
  & $\llbracket -\hat{N}_i p \nabla^{i}\bm{x}+U\rho\bm{v}\rrbracket=\bm{0}$ \\
\noalign{\smallskip}
$\delta t$
  & $\bm{\mathcal{E}}\cdot\bm{v}=0$
  & $\hat{n}_i p \nabla^{i}\bm{x}\cdot\bm{v}=0$
  & $\llbracket \hat{N}_i p \nabla^{i}\bm{x}\cdot\bm{v}-U\big(\tfrac{1}{2}\rho v^2+\rho e\big)\rrbracket=0$ \\
\noalign{\smallskip}
$\delta\eta^j$
  & $\bm{\mathcal{E}}\cdot\nabla_j\bm{x}=0$
  & $\hat{n}_i\big(\tfrac{1}{2}\rho v^2-\rho h\big)\delta^i_j=0$
  & $\llbracket -\hat{N}_i\big(\tfrac{1}{2}\rho v^2-\rho h\big)\delta^i_j -U\rho\bm{v}\cdot\nabla_j\bm{x}\rrbracket=0$ \\
\noalign{\smallskip}\hline\noalign{\smallskip}
\end{tabular}
\end{table}

Let us now note some further consequences of the pseudomomentum balance~\eqref{eq: pseudomomentum recast}. Taking a closed line integral over a material loop $\oint \mathrm{d}\eta^l$, with $\mathrm{d}\bm{l} = \nabla_l \bm{x} \mathrm{d}\eta^l$, we obtain  Kelvin's circulation theorem for $\Gamma$:
\begin{equation}
     d_t\Gamma = 0, \qquad \Gamma = \oint d_t \bm{x} \cdot \mathrm{d}\bm{l}.
    \label{kelvins circulation theorem}
\end{equation}
This result is also reported in  \cite{Cherepanov1977InvariantMechanics,Eckart1960VariationHydrodynamics,Padhye1996FluidSymmetry,Salmon1988HamiltonianMechanics,Singh2021Pseudomomentum:Consequences}. It is worth noting that, had the internal energy density been assumed to depend explicitly on position, the pseudomomentum balance would acquire a source term given by the material gradient of the corresponding internal energy density, but it would not change the conclusion in \eqref{kelvins circulation theorem}, as also mentioned in \cite{Singh2021Pseudomomentum:Consequences}. It is apparent that conservation of circulation $\Gamma$ is not merely a formal unguided manipulation of the governing equation but rather a deeper consequence of Noether's theorem as also mentioned in \cite[p.~131]{Badin2018VariationalDynamics} and \cite{Salmon1988HamiltonianMechanics}.\footnote{However, the remark made by Badin and Crisciani \cite{Badin2018VariationalDynamics} on the absence of such symmetry in elasticity is not correct as the elastic analog of such conservation is the $J$-integral in fracture mechanics \cite{Golebiewska-Herrmann1981OnMechanics}. In contrast, relabeling symmetry is genuinely absent in particle mechanics, where particles do not admit continuous relabeling, as correctly noted by Salmon \cite{Salmon1988HamiltonianMechanics}.}

A second consequence of the pseudomomentum balance~\eqref{eq: pseudomomentum recast} is obtained by integrating it with respect to time. Again, since $\nabla_j$ acting on a scalar involves no connection terms, it commutes with the time integration, and we obtain the Cauchy--Weber integral for compressible flow
\begin{equation}
    \Big(
    d_t\bm{x}\cdot \nabla_j\bm{x}
    -
    \big(d_t\bm{x}\cdot \nabla_j\bm{x}\big)\big|_{t=0}
    \Big)
    +\nabla_j\left[
    \int_0^t
    \left(
    \frac{p}{\rho}
    -\frac{1}{2} d_t\bm{x}\cdot d_t\bm{x}
    +e
    \right)\,\mathrm{d}t'
    \right]
    =0,
    \label{eq:cauchy-weber relation}
\end{equation}
where $d_t\bm{x}\cdot\nabla_j\bm{x}=v_j$ is the covariant component of the particle velocity along the material tangent $\nabla_j\bm{x}$ in the current configuration.
Equation~\eqref{eq:cauchy-weber relation}, which is also derived by Bennett \cite[Sec.~3.7]{Bennett2006LagrangianDynamics}, is thus the compressible analog of the Cauchy--Weber relation obtained in \cite{Singh2021Pseudomomentum:Consequences}  for incompressible flow.

A third consequence of \eqref{eq: bulk pseudomomentum balance} is obtained by projecting the pseudomomentum balance onto the velocity. Recall from the derivation of \eqref{eq: pseudomomentum recast} that, prior to applying \eqref{eq:dt_commute}, the balance reads $(d_t^2\bm{x})\cdot\nabla_j\bm{x}+\nabla_j h=0$, which is the material form of Euler's equation. Passing to the Eulerian description, the material acceleration is $d_t\bm{v}=\partial_t\bm{v}+\nabla(v^2/2)-\bm{v}\times\bm{\omega}$, with $\bm{\omega}=\nabla\times\bm{v}$ the vorticity, so that a kinetic term reappears from the convective part of the acceleration, carrying the sign opposite to the one it has in the pseudomomentum flux of \eqref{eq: pseudomomentum recast}. Projecting onto $\bm{v}$, we have  $\bm{v}\cdot(\bm{v}\times\bm{\omega})=0$, and we obtain
\begin{equation}
    \bm{v}\cdot\bigg[\partial_t \bm{v}+ \nabla\left(
    e+\frac{p}{\rho}+\frac{v^2}{2}
    \right)\bigg]=0.
    \label{eq:unsteady bernoulli eckart form}
\end{equation}
Equation~\eqref{eq:unsteady bernoulli eckart form} is the unsteady Bernoulli equation for compressible homentropic flow. Singh and Hanna \cite{Singh2021Pseudomomentum:Consequences} obtained the incompressible analog of \eqref{eq:unsteady bernoulli eckart form} from their energy balance, and noted that the streamline projection of the pseudomomentum balance carries the same content. For steady flow, it follows that the quantity $e+{p}/{\rho}+{v^2}/{2}$ is constant along a streamline.  Equation~\eqref{eq:unsteady bernoulli eckart form} can also be derived from the bulk energy balance in \eqref{eq:bulk_energy_balance}, using the material continuity equation \eqref{eq:continuity_material_frame}. Thus, in the compressible homentropic case, the streamline component of the pseudomomentum balance coincides with that of the energy balance, and both reduce to the Bernoulli equation.

Finally, if we did not project the balance $\partial_t\bm{v}+\nabla(v^2/2)-\bm{v}\times\bm{\omega}+\nabla h=\bm{0}$ onto $\bm{v}$ and the flow were irrotational, we could use a scalar potential \eqref{eq:velocity_potential} because  Kelvin's theorem \eqref{kelvins circulation theorem} guarantees that an initially irrotational flow remains such. In this case, the vorticity term now vanishes outright, and what remains of the balance is a pure gradient. Integrating it once recovers \eqref{eq:potential_interms_of_e}, now with an arbitrary function of integration:
\begin{equation}
    \partial_t\Psi
    -
    \frac{1}{2}\lvert\nabla\Psi\rvert^2
    -
    \frac{p}{\rho}
    -e
    =
    f(t),
    \label{cauchy-lagrange integral}
\end{equation}
where $f(t)$ is an arbitrary function of integration. Equation~\eqref{cauchy-lagrange integral} is termed the ``Cauchy--Lagrange integral'' by Sedov \cite[p.~156]{Sedov1972AMechanics}.

\subsection{Wave equations emerging from the Lagrangian variational principle}

We now repeat the perturbation scheme of Sec.~\ref{sec:perturbed_lagrangians} in the Lagrangian frame, taking the reference configuration to be the quiescent state $(\rho, p, e, \bm{v}) = (\bar{\rho}, \bar{p}, \bar{e}, \bm{0})$, so that $\bar\rho$, $\bar p = p(\bar\rho)$, and $\bar e = e(\bar\rho)$ are the constant $\rho_0$, $p_0$, and $e(\rho_0)$ of Sec.~\ref{sec:Eulerian}. The bookkeeping of Sec.~\ref{sec:perturbed_lagrangians} carries over: a Lagrangian density retained through order $n+1$ yields momentum and pseudomomentum balances accurate to order $n$; the energy balance is the exception, as we shall see.

\subsubsection{Perturbation expansion of the Lagrangian density in the material frame}
\label{sec:Lagrangian_perturbation}

For homentropic perturbations about the reference state $\rho=\bar\rho$, and anticipating which terms will contribute below, we expand the fields as
\begin{subequations}\label{eq:expansion_all}\begin{align}
    \rho
    &=
    \bar{\rho}
    + \rho^{(1)}
    + \rho^{(2)}
    +\text{h.o.t.},
    \label{eq:rho_expansion}\\
    d_t \bm{x}
    &= 
    \bm{0}
    + d_t \bm{x}^{(1)}
    + d_t \bm{x}^{(2)}
    +\text{h.o.t.},
    \label{eq:velocity_expansion}\\
    e(\rho)
    &=
    \bar e
    +
    \underbrace{\left.\frac{\partial e}{\partial \rho}\right|_{\rho=\bar{\rho}}\rho^{(1)}}_{=e^{(1)}}
    +
    \underbrace{\left[
    \left.\frac{\partial e}{\partial \rho}\right|_{\rho=\bar{\rho}}\rho^{(2)}
    +
    \frac{1}{2}
    \left.\frac{\partial^2 e}{\partial \rho^2}\right|_{\rho=\bar{\rho}}
    \big(\rho^{(1)}\big)^2
    \right]}_{=e^{(2)}}
    \label{eq:perturbative_expansions_constitutive_a}
    \\
    &\phantom{= \bar{e}} + \underbrace{\left[
    \left.\frac{\partial e}{\partial \rho}\right|_{\rho=\bar{\rho}}\rho^{(3)}
    +
    \left.\frac{\partial^2 e}{\partial \rho^2}\right|_{\rho=\bar{\rho}}
    \rho^{(1)}\rho^{(2)}
    + \frac{1}{6} \left.\frac{\partial^3 e}{\partial \rho^3}\right|_{\rho=\bar{\rho}}
    \big(\rho^{(1)}\big)^3
    \right]}_{=e^{(3)}}
    +
    \;\text{h.o.t.},
    \nonumber\\
    p(\rho)
    &=
    \bar p
    +
    \underbrace{\left.\frac{\partial p}{\partial \rho}\right|_{\rho=\bar{\rho}}\rho^{(1)}}_{=p^{(1)}}
    +
    \;\text{h.o.t.},
    \label{eq:perturbative_expansions_constitutive_b}
\end{align}\end{subequations}
with an implied asymptotic ordering, such as $\rho^{(1)} \ll \bar{\rho}$, etc. Here $\bm{x}^{(1)}$ is the first-order \emph{particle displacement} of a material element from its reference position, and $d_t\bm{x}^{(1)}$ the corresponding first-order particle velocity. These are the standard field variables of linear acoustics, expressed in the material description.

Before proceeding, we note the following expansion, which will be useful in the calculations that follow. Inverting the continuity constraint \eqref{eq:material_com} and expanding gives
\begin{subequations}\label{eq:J_expansions}\begin{equation}
    J^{-1}
    =
    \frac{\rho}{\bar{\rho}}
    =
    \frac{\bar{\rho} + \rho^{(1)} + \rho^{(2)} + \rho^{(3)} + \cdots}{\bar{\rho}}
    =
    1 + \frac{\rho^{(1)}}{\bar{\rho}} + \frac{\rho^{(2)}}{\bar{\rho}} + \frac{\rho^{(3)}}{\bar{\rho}} + \text{h.o.t.}
    \label{eq:J_inverse_expansion} 
\end{equation}
Following Bouchet et al.~\cite{Bouchet1995PerturbativeInstability}, the Jacobian may also be expanded in a perturbation series, $J=1+J^{(1)}+J^{(2)}+J^{(3)}+\cdots$, and its inverse follows directly from the identity $J\,J^{-1}=1$:
\begin{equation}
    J^{-1}
    =
    {1-J^{(1)}}
    +
    \left[(J^{(1)})^2-J^{(2)}\right]
    + 
    \left[-(J^{(1)})^3 + 2J^{(1)}J^{(2)} - J^{(3)}\right]
    +\text{h.o.t.}
    \label{eq:J_inverse_expansion_b}
\end{equation}\end{subequations}
By comparing \eqref{eq:J_inverse_expansion} with \eqref{eq:J_inverse_expansion_b} order by order, one obtains the following relations among the successive terms in the density perturbation:
\begin{subequations}\label{eq:density_perturbation_orders}
\begin{align}
    \rho^{(1)}
    &=
    -\bar\rho\,J^{(1)},
    \label{eq:density_perturbation_orders_a}
    \\
    \rho^{(2)}
    &= 
    \bar\rho \left[(J^{(1)})^2 - J^{(2)} \right]=\frac{\big(\rho^{(1)}\big)^2}{\bar\rho}-\bar \rho J^{(2)},
    \label{eq:density_perturbation_orders_b} \\
    \rho^{(3)}
    &= \bar\rho \left[-(J^{(1)})^3 + 2J^{(1)}J^{(2)} - J^{(3)}\right].
    \label{eq:density_perturbation_orders_c}
\end{align}
\end{subequations}
We also note the identities reported in \cite{Bouchet1995PerturbativeInstability}:
\begin{subequations}\label{eq:J1_J2_identities}
    \begin{align}
        J^{(1)} &=\nabla\cdot\bm{x}^{(1)}, \\
        J^{(2)} &= \nabla\cdot\bm{x}^{(2)} + \frac{1}{2} \left[ \big(\nabla\cdot\bm{x}^{(1)}\big)^2 -\nabla\bm{x}^{(1)}:\big(\nabla\bm{x}^{(1)}\big)^\top \right].
    \end{align}
\end{subequations}
The particle position field may be expanded as $\bm{x} = \bm{x}^{(0)} + \bm{x}^{(1)} + \bm{x}^{(2)} + \cdots$, where $\bm{x}^{(0)}=\bar{\bm{x}}$ is the prescribed reference position, so that $d_t\bm{x}^{(0)}=\delta\bm{x}^{(0)}=\bm{0}$.

Substituting the perturbation expansions of the internal energy from \eqref{eq:perturbative_expansions_constitutive_a} into the Lagrangian density \eqref{eq:compressible lagrangian}, and truncating to third order in the perturbation expansion yields the perturbative Lagrangian density
\begin{multline}
    \bar{\mathcal{L}}_3
    = \underbrace{-\bar{\rho} \left.\frac{\partial e}{\partial \rho}\right|_{\rho=\bar{\rho}} \rho^{(1)}}_{\bar{\mathcal{L}}^{(1)}} + \underbrace{\frac{1}{2} \bar{\rho} \, d_t \bm{x}^{(1)} \cdot d_t \bm{x}^{(1)} - \bar{\rho} \left[ \left.\frac{\partial e}{\partial \rho}\right|_{\rho=\bar{\rho}} \rho^{(2)} + \frac{1}{2} \left.\frac{\partial^2 e}{\partial \rho^2}\right|_{\rho=\bar{\rho}} \big(\rho^{(1)}\big)^2 \right]}_{\bar{\mathcal{L}}^{(2)}} \\
    + \underbrace{\bar{\rho}  d_t \bm{x}^{(1)} \cdot d_t \bm{x}^{(2)} - \bar{\rho} \left[ \left.\frac{\partial e}{\partial \rho}\right|_{\rho=\bar{\rho}}\rho^{(3)} + \left.\frac{\partial^2 e}{\partial \rho^2}\right|_{\rho=\bar{\rho}} \rho^{(1)} \rho^{(2)} + \frac{1}{6} \left.\frac{\partial^3 e}{\partial \rho^3}\right|_{\rho=\bar{\rho}} \big(\rho^{(1)}\big)^3 \right]}_{\bar{\mathcal{L}}^{(3)}} ,
    \label{eq:cubic_lagrangian_lagrangian}
\end{multline}
where the zeroth-order contribution $-\bar{\rho}\,\bar{e}$ has been discarded from \eqref{eq:cubic_lagrangian_lagrangian} as an additive constant.

\runinhead{Further assumptions and reductions.} 
Equation~\eqref{eq:cubic_lagrangian_lagrangian} is the most general third-order truncation of the Lagrangian density in the material frame. 
Note that the first term in \eqref{eq:cubic_lagrangian_lagrangian} is \emph{first} order in the perturbation; its presence is consistent with truncation at third order, which retains all terms through that order, and it does not affect the bulk dynamics. To see this, eliminate $\rho^{(2)}$ from the term $(\partial e/\partial\rho)|_{\rho=\bar{\rho}} \,\rho^{(2)}$  using \eqref{eq:density_perturbation_orders_b}, and $\rho^{(3)}$ using \eqref{eq:density_perturbation_orders_c} and recall $\bar{p}=\bar{\rho}^{2}(\partial e/\partial\rho)|_{\rho=\bar{\rho}}$ from \eqref{eq:d_thermo_de} together with \eqref{eq:density_perturbation_orders_a}. Then, \eqref{eq:cubic_lagrangian_lagrangian} turns into
\begin{equation}
    \begin{split}
    \bar{\mathcal{L}}_3
    &=
    \frac{1}{2}\bar{\rho}
    d_t\bm{x}^{(1)}
    \cdot
    d_t\bm{x}^{(1)}
    +
    \bar{\rho}
    d_t\bm{x}^{(1)}
    \cdot
    d_t\bm{x}^{(2)}
    -
    \big(\rho^{(1)}\big)^2\underbrace{
    \left[
    \left.
    \frac{\partial e}{\partial\rho}
    \right|_{\rho=\bar{\rho}}
    +
    \frac{\bar{\rho}}{2}
    \left.
    \frac{\partial^2 e}{\partial\rho^2}
    \right|_{\rho=\bar{\rho}}
    \right]}_{c_0^2/ (2\bar\rho)}
    \\
    &\quad -
    \rho^{(1)}\rho^{(2)}
    \underbrace{\left[
    2\left.
    \frac{\partial e}{\partial\rho}
    \right|_{\rho=\bar{\rho}}
    +
    \bar{\rho}
    \left.
    \frac{\partial^2 e}{\partial\rho^2}
    \right|_{\rho=\bar{\rho}}
    \right]}_{c_0^2 / \bar\rho}
    +
    \underbrace{\left[
    \frac{1}{\bar{\rho}}
    \left.
    \frac{\partial e}{\partial\rho}
    \right|_{\rho=\bar{\rho}}
    -
    \frac{\bar{\rho}}{6}
    \left.
    \frac{\partial^3 e}{\partial\rho^3}
    \right|_{\rho=\bar{\rho}}
    \right]}_{\frac{1}{6\bar{\rho}^2}\left( 4c_0^2 - \bar\rho{\left.\frac{\partial^2 p}{\partial \rho^2}\right|_{\rho=\bar{\rho}}}\right)}
    \big(\rho^{(1)}\big)^3
    \\
    &\quad +
    \underbrace{
    \bar{p}
    \left(
    J^{(1)}+J^{(2)}+J^{(3)}
    \right)
    }_{\text{boundary term}} .
    \end{split}
    \label{eq:cubic-lagrangian-collected}
\end{equation}

The underbrace in \eqref{eq:cubic-lagrangian-collected} already anticipates the first of two assumptions we shall make in reducing the Lagrangian density to a form from which a weakly nonlinear acoustic wave equation can be derived. First, after some algebra, it can be shown that $J^{(1)}$, $J^{(2)}$, and $J^{(3)}$ are divergences~\cite{Bouchet1995PerturbativeInstability}, and $\bar{p}$ is constant. But, in the framework of Sec.~\ref{action symmetries and implied balance}, a total divergence does \emph{not} simply vanish: it produces terms in $Q$ and $\mathcal{J}^{i}$ in \eqref{eq:schematic_deltaA}, and hence the boundary and jump conditions. Calling $\bar{p} (J^{(1)}+J^{(2)}+J^{(3)})$ a ``boundary term'' and discarding it therefore amounts to restricting our attention to the bulk dynamics.

With this in mind, \eqref{eq:cubic-lagrangian-collected} simplifies to
\begin{multline}
    \bar{\mathcal{L}}_3
    =
    \frac{1}{2} \bar{\rho}
    d_t\bm{x}^{(1)}
    \cdot
    d_t\bm{x}^{(1)}
    +
    \bar{\rho} 
    d_t\bm{x}^{(1)}
    \cdot
    d_t\bm{x}^{(2)}
    -
    \frac{c_0^2}{2\bar{\rho}}
    \big(\rho^{(1)}\big)^2 \\
    -
    \frac{c_0^2}{\bar{\rho}}
    \rho^{(1)}\rho^{(2)}
    +
    \frac{4c_0^2-\bar{\rho}\left.\frac{\partial^2 p}{\partial \rho^2}\right|_{\rho=\bar{\rho}}}
    {6\bar{\rho}^{2}}
    \big(\rho^{(1)}\big)^3 .
    \label{eq:cubic lagrangian simplified}
\end{multline}
Using the identities \eqref{eq:J1_J2_identities}, the second-order perturbation of density $\rho^{(2)}$ can be written as
\begin{equation}
    \rho^{(2)}
    =
    -\bar{\rho}
    \nabla\cdot\bm{x}^{(2)}
    +
    \frac{\bar{\rho}}{2}
    \left[
    \big(
    \nabla\cdot\bm{x}^{(1)}
    \big)^2
    +
    \nabla\bm{x}^{(1)}
    :
    \big(
    \nabla\bm{x}^{(1)}
    \big)^\top
    \right].
    \label{eq:rho second order}
\end{equation}
Substituting $\rho^{(2)}$ from \eqref{eq:rho second order} into \eqref{eq:cubic lagrangian simplified} and using the linearized constitutive relation $p^{(1)} = c_0^2\rho^{(1)}$ from \eqref{eq:perturbative_expansions_constitutive_b}, the reduced Lagrangian density becomes
\begin{multline}
    \bar{\mathcal{L}}_3
    = \frac{1}{2} \bar{\rho} d_t\bm{x}^{(1)} \cdot d_t\bm{x}^{(1)} - \frac{c_0^2}{2\bar{\rho}} \big(\rho^{(1)}\big)^2 + \overbrace{\bar{\rho} d_t\bm{x}^{(1)} \cdot d_t\bm{x}^{(2)} + p^{(1)} \nabla \cdot \bm{x}^{(2)}}^{\text{terms containing } \bm{x}^{(2)}} \\
    - \frac{c_0^2}{2} \rho^{(1)} \nabla \bm{x}^{(1)} : \big(\nabla \bm{x}^{(1)}\big)^\top 
    + 
    \frac{c_0^2 - \bar{\rho} \left.\frac{\partial^2 p}{\partial \rho^2}\right|_{\rho=\bar{\rho}}}{6\bar{\rho}^2}
    \big(\rho^{(1)}\big)^3 .
\label{eq: cubic lagrangian simplify}
\end{multline}

Second, the overbraced terms in \eqref{eq: cubic lagrangian simplify} still carry the second-order displacement $\bm{x}^{(2)}$, and, unlike the divergence discarded above, they are not divergences at all. Integrating them by parts, they multiply $\bar\rho d_t^2\bm{x}^{(1)} + \nabla_i(p^{(1)}\nabla^i\bm{x})$, the linearization of the bulk momentum balance \eqref{eq: bulk momentum}, derived below as \eqref{eq:linearized_momentum_balance}, and hence, on solutions of that balance, reduce to spatial and temporal boundary terms, contributing nothing to the bulk dynamics (invoking the equation of motion rather than an identity).

Recalling the nonlinearity parameter \eqref{eq:BoverA}, the coefficient of $(\rho^{(1)})^3$ in \eqref{eq: cubic lagrangian simplify} becomes $c_0^{2}(1-B/A)/6\bar{\rho}^{2}$, and dropping the $\bm{x}^{(2)}$ terms, \eqref{eq: cubic lagrangian simplify} reduces to
\begin{multline}
    \bar{\mathcal{L}}_3
    = \frac{1}{2} \bar{\rho} d_t\bm{x}^{(1)} \cdot d_t\bm{x}^{(1)} - \frac{c_0^2}{2\bar{\rho}} \big(\rho^{(1)}\big)^2\\
    - \frac{c_0^2}{2} \rho^{(1)} \nabla \bm{x}^{(1)} : \big(\nabla \bm{x}^{(1)}\big)^\top 
    + \frac{c_0^2}{6\bar{\rho}^2}\left(1 - \frac{B}{A}\right) \big(\rho^{(1)}\big)^3 .
\label{eq: cubic lagrangian final form}
\end{multline}

\subsubsection{Linear acoustics in the material frame}
\label{sec:Lagrangian_linear}

In this section, we derive the balance laws of linear acoustics in two complementary ways. First, we linearize the exact momentum, energy, and pseudomomentum balances, \eqref{eq: bulk momentum}, \eqref{eq: energy balance final form}, and \eqref{eq: pseudomomentum recast}, obtained in Sec.~\ref{action symmetries and implied balance}. Second, we truncate \eqref{eq: cubic lagrangian final form} at quadratic order and apply the variational procedures developed in Secs.~\ref{background and useful preliminaries} and \ref{action symmetries and implied balance} to derive the corresponding linearized balance laws.

We then compare the two procedures: the momentum and pseudomomentum balances come out identical, while time-translation symmetry yields the acoustic energy balance one order higher. Introducing the approximation directly at the level of the Lagrangian density thus does not destroy the relevant variational symmetries or their associated conservation laws.

Substituting the expansion \eqref{eq:expansion_all} into the bulk momentum balance~\eqref{eq: bulk momentum} and collecting the perturbation terms yields the linearized bulk momentum balance:
\begin{subequations}\begin{equation}
    \bar{\rho} d_t^2 \bm{x}^{(1)}
    + \nabla_i  \left( p^{(1)} \nabla^i \bm{x} \right)
    = \bm{0}.
    \label{eq:linearized_momentum_balance}
\end{equation}
Note that the zeroth-order pressure does not contribute to \eqref{eq:linearized_momentum_balance} since $\bar p$ is uniform, so $\nabla_i(\bar p \nabla^i\bm{x}) = \bar p \nabla^2\bm{x} = \bm{0}$ identically by \eqref{eq:flatness_identity}, and, $p^{(1)}$ being already first order, $\nabla^i\bm{x}$ in the surviving term may be evaluated on the reference configuration.
Then, substituting the expansion \eqref{eq:expansion_all} into the bulk pseudomomentum balance~\eqref{eq: pseudomomentum recast}, and retaining only terms linear in the perturbations gives
\begin{equation}
    d_t \left(
        d_t \bm{x}^{(1)} \cdot \nabla_j \bm{x}
    \right)
    + \nabla_j \left(
        \frac{p^{(1)}}{\bar{\rho}}
    \right)
    =0.
    \label{eq:linearized_pseudomomentum_balance}
\end{equation}
Finally, linearizing the bulk energy balance~\eqref{eq: energy balance final form} leads to
\begin{equation}
    \bar{\rho} d_t e^{(1)}
    + \bar{p} \nabla_i \left(
        \nabla^i \bm{x} \cdot d_t \bm{x}^{(1)}
    \right)
    = 0.
    \label{eq:linearized_energy_balance}
\end{equation}\end{subequations}
Equations \eqref{eq:linearized_momentum_balance},
\eqref{eq:linearized_pseudomomentum_balance}, and \eqref{eq:linearized_energy_balance} are the governing equations of configurational linear acoustics. Consistent
with Sec.~\ref{sec:consequences}, they are not independent, and each may be used
in place of the others.  Here and below, the linearized \eqref{eq:continuity_material_frame} reads $d_t\rho^{(1)} = -\bar{\rho}\,\nabla^i\bm{x}\cdot\nabla_i(d_t\bm{x}^{(1)})$. Note that \eqref{eq:linearized_energy_balance} is $\bar p$ times this relation and so holds identically.

Now we present an alternative approach to obtaining the linearized balances using the framework developed in Secs.~\ref{background and useful preliminaries} and \ref{action symmetries and implied balance}. Truncating \eqref{eq: cubic lagrangian final form} at quadratic order and using $p^{(1)} = c_0^2\rho^{(1)}$ from
\eqref{eq:perturbative_expansions_constitutive_b}, we obtain
\begin{equation}
    \bar{\mathcal{L}}_2
    =
    \frac{1}{2}\bar{\rho}\, d_t\bm{x}^{(1)}\cdot d_t\bm{x}^{(1)}
    -\frac{\big(p^{(1)}\big)^{2}}{2\bar{\rho}c_0^{2}},
    \label{eq:L2_material}
\end{equation}
Using \eqref{eq:var_action_standard_format} and collecting the terms conjugate to $\delta\bm{x}$, $\delta t$, and $\delta\eta^j$ in turn, as in Sec.~\ref{action symmetries and implied balance}, we obtain the bulk momentum, pseudomomentum, and energy balances:
\begin{subequations}\begin{align}
    \bar{\rho} d_t^2 \bm{x}^{(1)}
    + \nabla_i \left( p^{(1)} \nabla^i \bm{x} \right)
    &= \bm{0},
    \label{eq:bulk_momentum_balance_v2}\\
    d_t \left( d_t \bm{x}^{(1)} \cdot \nabla_j \bm{x} \right)
    + \nabla_j \left(\frac{p^{(1)}}{ \bar \rho} 
    \right)
    &= 0,
    \label{eq:bulk_pseudomomentum_balance_v2}\\
    d_t\left[
    \frac{1}{2}\bar{\rho}\, d_t\bm{x}^{(1)}\cdot d_t\bm{x}^{(1)}
    +
    \frac{\big(p^{(1)}\big)^2}{2\bar{\rho}c_0^2}
    \right]
    +
    \nabla_i\big(p^{(1)}\nabla^i\bm{x}\cdot d_t\bm{x}^{(1)}\big)
    &= 0 .
    \label{eq:bulk_energy_balance_v2}
\end{align}\end{subequations}
Here \eqref{eq:bulk_momentum_balance_v2} is the statement $\bm{\mathcal{E}}^{(1)}=\bm{0}$, with $-\bm{\mathcal{E}}^{(1)} \equiv \bar{\rho}\, d_t^2\bm{x}^{(1)} + \nabla_i\big( p^{(1)}\nabla^i\bm{x}\big)$ the first-order term, while \eqref{eq:bulk_pseudomomentum_balance_v2} and \eqref{eq:bulk_energy_balance_v2} are its projections onto $\nabla_j\bm{x}$ and $d_t\bm{x}^{(1)}$, the latter rearranged by the product rule and the linearized \eqref{eq:continuity_material_frame} above. Indeed, \eqref{eq:bulk_momentum_balance_v2} and \eqref{eq:bulk_pseudomomentum_balance_v2} coincide exactly with \eqref{eq:linearized_momentum_balance} and \eqref{eq:linearized_pseudomomentum_balance}, respectively. The energy balance enters differently: \eqref{eq:bulk_energy_balance_v2} is second order because $d_t\bm{x}^{(1)}$ is itself first order, whereas $\nabla_j\bm{x}$ is not. Its first-order counterpart \eqref{eq:linearized_energy_balance} is instead the linearization of \eqref{eq:internal_energy_rate}, following from the constraint and the equation of state rather than from the symmetry. Pierce \cite[Eq.~(1.11.9)]{Pierce2019Acoustics} derives it in the Eulerian setting; see also \cite[Sec.~6.2]{Morse1968TheoreticalAcoustics}.

This calculation demonstrates that one may linearize the fields and construct the corresponding approximate Lagrangian rather than directly linearizing the governing PDEs. We note that this calculation provides another demonstration of Salmon's observation that an approximation made at the level of the Lagrangian density does not destroy the corresponding symmetry or conservation properties
\cite{Salmon1988HamiltonianMechanics}.

Either the linearized momentum balance~\eqref{eq:linearized_momentum_balance} or the linearized pseudomomentum balance~\eqref{eq:linearized_pseudomomentum_balance} suffices to eliminate the particle displacement in favor of a
single scalar field. For a space-filling fluid in flat Euclidean space, the identities
\eqref{eq:tangent_identity} and \eqref{eq:flatness_identity} hold exactly. Taking the divergence of the linearized momentum balance~\eqref{eq:linearized_momentum_balance}, and using $\rho^{(1)}=-\bar{\rho}\nabla\cdot\bm{x}^{(1)}$ from
\eqref{eq:density_perturbation_orders_a} together with $p^{(1)}=c_0^{2}\rho^{(1)}$, the displacement is eliminated and the familiar
linear acoustic wave equation is recovered:
\begin{equation}
    d_t^2 \rho^{(1)} - c_0^2 \nabla^2 \rho^{(1)} = 0,
    \label{eq:density wave equation from momentum balance}
\end{equation}
where
$c_0^2=({\partial p}/{\partial \rho})|_{\rho=\bar{\rho}}$ as before. 

At leading order, the material and Eulerian time derivatives coincide, $d_t=\partial_t$, so \eqref{eq:density wave equation from momentum balance} may just as well be read as a PDE in the Eulerian sense. Thus,  \eqref{eq:density wave equation from momentum balance} is again the acoustic wave equation~\eqref{eq:morse_ingard_wave_equation}  but in terms of the density perturbation $\rho^{(1)}$, also reported in textbooks (e.g., \cite{Blackstock2000FundamentalsAcoustics,Gonzalez2001AMechanics}). 
Similarly, taking $d_t$ of \eqref{eq:linearized_pseudomomentum_balance} and eliminating the displacement with the linearized \eqref{eq:continuity_material_frame} also yields \eqref{eq:density wave equation from momentum balance}. In flat Euclidean space, the linearized momentum and pseudomomentum balances both yield the classical linear wave equation.

We now return to $\bar{\mathcal{L}}_2$ in \eqref{eq:L2_material}, the familiar ``kinetic minus potential energy'' Lagrangian density of linear acoustics. It is precisely the quantity termed ``the second-order Lagrangian density'' by Hamilton and Morfey~\cite[Eq.~(3.38)]{Hamilton2024ModelEquations}, which they obtain as an algebraic grouping in the course of assembling the second-order wave equation (rather than from a variational principle). Not coincidentally, the conserved density in \eqref{eq:bulk_energy_balance_v2} is the acoustic energy density (kinetic \emph{plus} potential) and its flux ($p^{(1)}\nabla^i\bm{x}\cdot d_t\bm{x}^{(1)}$) is the acoustic intensity \cite{Morse1968TheoreticalAcoustics,Pierce2019Acoustics,Landau1987FluidMechanics}. It is the Hamiltonian conjugate to \eqref{eq:L2_material}.

It is equally instructive to compare \eqref{eq:L2_material} with the quadratic Lagrangian density \eqref{eq:L_linearized_from_final} obtained in the Eulerian frame. Restricting now, as in Sec.~\ref{sec:Eulerian}, to an initially irrotational flow, at leading order the material velocity is the Eulerian one, $d_t\bm{x}^{(1)}=\bm{v}^{(1)}=-\nabla\Psi$, while linearizing the Cauchy--Lagrange integral \eqref{cauchy-lagrange integral} gives $\partial_t\Psi=h^{(1)}=p^{(1)}/\bar{\rho}$. Substituting both into \eqref{eq:L2_material} yields
\begin{equation}
    \bar{\mathcal{L}}_2 = \tfrac12\bar\rho|\nabla\Psi|^2
  - \frac{\bar\rho}{2c_0^2}\big(\partial_t\Psi\big)^2 = -\bar\rho \mathcal{L}_2 ,
  \qquad \bar\rho = \rho_0 .
    \label{eq:L2_frames_agree}
\end{equation}
The second-order Lagrangian densities obtained in the two frames therefore coincide up to rescaling.\footnote{Note that the minus sign comes from passing from the particle displacement to the velocity potential, which interchanges the roles of the kinetic and compressional terms.} The two variational principles are not merely consistent; they are the same principle (when the flows are both irrotational) expressed in different variables!

\subsubsection{Weakly nonlinear acoustics in the material frame}
\label{sec:Lagrangian_weakly_nonlinear}

We now return to the cubic Lagrangian density \eqref{eq: cubic lagrangian final form}, focusing on its one-dimensional (1D) reduction. Let us denote $\bm{x}^{(1)}(\eta,t) = \xi(\eta,t)\hat{\bm{e}}$, where $\hat{\bm{e}}$ is the unit vector along the direction of propagation and $\eta$ is the material coordinate along it. Then the first-order velocity becomes $d_t\bm{x}^{(1)} = \xi_t \hat{\bm{e}}$, and $\nabla \bm{x}^{(1)} = \xi_{\eta} \, \hat{\bm{e}} \otimes \hat{\bm{e}}$. Using the 1D notation, the first order perturbation in density $\rho^{(1)}$ in \eqref{eq:density_perturbation_orders_a} can be written as $\rho^{(1)}=-\bar\rho\xi_{\eta}$. Using these results, \eqref{eq: cubic lagrangian final form} becomes
\begin{equation}
    \bar{\mathcal{L}}_{3,\mathrm{1D}}=\frac{1}{2} \bar{\rho} \xi_t^2 - \frac{1}{2} \bar{\rho} c_0^2 \xi_\eta^2 + \frac{1}{3}\left(1 + \frac{B}{2A}\right) \bar{\rho} c_0^2 \xi_\eta^3.    
    \label{eq:1D cubic lagrangian}
\end{equation}
This expression features $1+B/(2A)$, the fluid's coefficient of nonlinearity, whereas its Eulerian counterpart in \eqref{eq:cubic_truncated_lagrangian} featured $1-B/A$. The two need not agree: \eqref{eq:cubic_truncated_lagrangian} is a functional in $\Psi$ and \eqref{eq:1D cubic lagrangian} is in $\xi$, so only the wave equations they generate are comparable.

Unlike \eqref{eq:L2_material}, which we varied over trajectories, \eqref{eq:1D cubic lagrangian} depends on the time and spatial derivatives of a single field, so we may treat it as a field Lagrangian and apply the Euler--Lagrange operator \eqref{eq:EL operator} directly, giving
\begin{equation}
    \xi_{tt} = c_0^2 \left[ 1 - 2\left(1+\frac{B}{2A}\right) \xi_\eta \right] \xi_{\eta\eta}.
    \label{lamb from cubic lagrangian 1d}
\end{equation}
Equation~\eqref{lamb from cubic lagrangian 1d} holds for any nondissipative homentropic fluid. Specializing now to a perfect gas, \eqref{eq:isentropic_rels} gives $B/A = \gamma-1$, so $2[1+B/(2A)] = \gamma+1$, giving
\begin{equation}
    \xi_{tt} = c_0^2 \left[ 1 - (\gamma + 1) \xi_\eta \right] \xi_{\eta\eta}.
    \label{eq:lamb perfect gas}
\end{equation}
We interpret \eqref{lamb from cubic lagrangian 1d} as a material version of the weakly nonlinear wave equation \eqref{eq: weakly nonlinear equation from L3}, whose perfect-gas form \eqref{eq:lamb perfect gas} was also discussed by Lamb \cite[\S63, Eq.~(25)]{Lamb1960TheSound}. Peierls \cite[Eq.~(9.3)]{Peierls1985} obtained a dispersive generalization from a Lagrangian with weakly anharmonic potential energy; his equation reduces to \eqref{lamb from cubic lagrangian 1d} when the dispersion coefficient vanishes, with his anharmonic coefficient playing the role of our $2(1 + B/(2A))$. We also stress a distinction: Lamb reduces the exact material-frame equation of motion\footnote{That is, $\xi_{tt} = c_0^2(1+\xi_\eta)^{-(\gamma+1)}\xi_{\eta\eta}$, often attributed to Earnshaw; see \cite{Blackstock2024History1750s1930s}.} \cite[\S63, Eq.~(9)]{Lamb1960TheSound}, whereas \eqref{eq:lamb perfect gas} follows from truncating the Lagrangian. Their agreement is the nonlinear counterpart of the linear equivalence established in Sec.~\ref{sec:Lagrangian_linear} above, and a further instance of Salmon's observation that an approximation made at the level of the Lagrangian density does not destroy the corresponding symmetry or conservation properties \cite{Salmon1988HamiltonianMechanics}. Finally, we leave an exercise for the reader.

\begin{svgraybox}
\begin{exercise}
Repeat the derivation of Sec.~\ref{sec:Lagrangian_perturbation} using the single displacement field $\bm{u}=\bm{x}-\bar{\bm{x}}$ in place of the order-by-order expansion $\bm{x}=\bar{\bm{x}}+\bm{x}^{(1)}+\bm{x}^{(2)}+\cdots$. Since $\bar{\bm{x}}$ is time independent, show first that the kinetic energy density $\tfrac{1}{2}\bar\rho d_t\bm{u}\cdot d_t\bm{u}$ is \emph{exactly} quadratic, with no expansion required, so that the cross terms encountered above never appear. All of the nonlinearity then resides in the stored energy
\begin{equation*}
    W(J) = \bar\rho\,e\big(\bar\rho J^{-1}\big),
    \qquad J = \det\big(\bm{I}+\bar\nabla\bm{u}\big),
\end{equation*}
which depends on the deformation only through the volume ratio $J$: a barotropic fluid is a hyperelastic material \cite{Truesdell2004TheMechanics}! Show that
\begin{equation*}
    W'(J) = -p,
    \qquad
    W''(1) = \bar\rho c_0^{2},
    \qquad
    W'''(1) = -2\bar\rho c_0^{2}\left(1+\frac{B}{2A}\right),
\end{equation*}
so that the coefficient appearing in \eqref{lamb from cubic lagrangian 1d} is nothing but $-W'''(1)/(\bar\rho c_0^{2})$, the third derivative of the stored energy at the reference state. Expanding $W$ to cubic order in $\bar\nabla\bm{u}$ and varying should reproduce \eqref{eq: cubic lagrangian final form}. Why must the two routes agree, and what became of the terms that had to be removed using the linearized momentum balance in Sec.~\ref{sec:Lagrangian_perturbation}?
\end{exercise}
\end{svgraybox}

\section{Conclusion}
\label{sec:conclusion}

In this chapter, we have addressed the somewhat elusive origin of the Lagrangian densities that appear throughout the acoustics literature. We showed that such Lagrangians need not be introduced by guesswork, but may instead be derived systematically from the primitive Lagrangian density of classical continuum mechanics. We then developed a perturbative treatment starting from the general variational principle and showed how the perturbative expansion of the Lagrangian density reproduces, at linear order, the same acoustic wave equation as that obtained by linearizing the balance laws and subsequently eliminating auxiliary variables. Keeping more terms in the perturbation expansion (specifically, to cubic order) yields a weakly nonlinear equation reported by Rasmussen et al.~\cite[Eq.~(7)]{Rasmussen2008AnalyticalDissipation}, wherein it was obtained by introducing approximations directly at the level of the governing PDEs. 

Having completed the discussion in the Eulerian setting, we next turned to the Lagrangian description in order to examine whether the relabeling symmetry, so central in elasticity and configurational mechanics, yields an equally meaningful consequence in acoustics. As discussed before \cite{Singh2021Pseudomomentum:Consequences}, this formulation naturally leads to the pseudomomentum balance and, consequently, to Kelvin's circulation theorem, as well as various integrals named after Cauchy, Lagrange, Weber, and Bernoulli. Unsurprisingly, in the absence of sources of pseudomomentum, the streamline component of the pseudomomentum balance coincides with that of the energy balance. We carried out a perturbative treatment of the Lagrangian density in the configurational context, provided a coherent understanding of the ``second-order Lagrangian density'' quoted in the literature, and recovered the equations of linear acoustics. That the momentum and pseudomomentum balances coincide exactly with those obtained by expanding the balance laws themselves, while the time-translation symmetry yields instead the second-order acoustic energy balance, indicates that perturbation expansion at the level of the Lagrangian density preserves, rather than destroys, the underlying variational structure, and may be a fruitful approach for obtaining principled approximations in acoustics. Carrying the same expansion to cubic order, we obtained a weakly nonlinear Lagrangian density in the material frame, whose one-dimensional reduction yields a weakly nonlinear wave equation for any barotropic fluid, with the nonlinearity carried by $B/A$, reducing for a perfect gas to the equation reported by Lamb \cite[\S63, Eq.~(25)]{Lamb1960TheSound}. Notably, our equation follows from truncating the Lagrangian density and then varying it, rather than truncating the equation of motion.

A thread we uncovered but did not pull on concerns the relation between the pseudomomentum balance of Sec.~\ref{sec:configurational} and the momentum--pseudomomentum distinction that Peierls placed at the center of the Abraham--Minkowski problem \cite{Peierls1976TheMedium}. Peierls treated light and sound on the same footing \cite[\S4.2]{Peierls1979SurprisesPhysics}, and suggested that ``[p]seudomomentum is also applicable to the problem of sound pressure,'' yet he explicitly chose not to pursue the point (``I shall not discuss this here'') \cite[p.~248]{Peierls1985}. To the best of our knowledge, no one has followed up on this point. Meanwhile, the subject of the acoustic radiation pressure ``has been associated with a lot of confusion and controversy'' \cite{Lee1993AcousticPressure}: the force on an absorbing target is fixed by the pseudomomentum flux, while the momentum actually carried by a wavepacket depends on the mean-flow response \cite{Andrews1978AnFlow,McIntyre1981OnMyth,Stone2000AcousticMedium}. Whether the radiation force can be recovered as a path-independent integral of the pseudomomentum flux remains an open question.

What we hope the reader takes away from our discussion here is that the unapproximated governing equations of nondissipative flows of compressible fluids have an elegant variational structure, which ought to be appreciated and explored ``as is.'' We have focused on nondissipative, homentropic fluids; irrotationality and the perfect-gas closure enter only as further specializations to the acoustics context, where they are needed. We do not wish to get into thorny issues of comparing various weakly nonlinear models \cite{Christov2007ModelingPropagation,Jordan2009WeaklyAnalysis,Jordan2012OnFluids} or classifying them in hierarchies \cite{Jordan2024AnSystem,Jordan2025Updating2018}. The two frames differ in one respect worth emphasizing, namely that without introducing the machinery of Clebsch variables, irrotationality is forced on the Eulerian principle by the mass constraint needed to make that description work, while the Lagrangian calculation never imposes it, yielding Kelvin's theorem \eqref{kelvins circulation theorem} instead.


\begin{acknowledgement}
We dedicate this work to Dr.\ P.\ M.\ Jordan on the occasion of his retirement, and in celebration of his contributions to acoustics in fluids and in porous media, as well as to second sound theories. We also thank Profs.\ Ada Amendola, Vanja Nikoli\'{c}, Brian Straughan, and Vittorio Zampoli for their kind invitation to contribute to this volume and for their efforts in co-organizing it. 
We thank Profs.\ Harmeet Singh and James Hanna for the fruitful discussion on configurational mechanics \cite{Singh2021Pseudomomentum:Consequences} and for pushing us to explore the consequences of pseudomomentum balances in fluid mechanics. 
I.C.C.\ further acknowledges many insightful discussions with P.M.J.\ on weakly nonlinear acoustics equations, including initial forays into variational acoustics \cite{Christov2015CorrectionsGases}.
The authors acknowledge partial support by the U.S.\ National Science Foundation under grant No.~CMMI-2245343 during the completion of this work.
\end{acknowledgement}

\ethics{Competing Interests}{The authors have no conflicts of interest to declare that are relevant to the content of this chapter.}


\footnotesize{
\bibliographystyle{spbasic-mod}
\bibliography{ref_cleaned}
}

\end{document}